\documentclass[hyper,twoside]{JHEP3}

\pdfoutput=1
\usepackage{multicol}
\usepackage{amssymb}
\usepackage{amsmath}
\usepackage{graphicx}

\def\urltilda{\kern -.15em\lower .7ex\hbox{\~{}}\kern .04em}
\def\deg{^{\circ}}
\newcommand{\Fermi}{\emph{Fermi} }
\newcommand{\FLAT}{\emph{Fermi}-LAT }
\title{Direct Constraints on Minimal Supersymmetry from Fermi-LAT Observations of the Dwarf Galaxy Segue 1}

\author{Pat Scott\\
	Oskar Klein Centre for Cosmoparticle Physics and\\
        Department of Physics, Stockholm University\\
        AlbaNova, SE-10691 Stockholm, Sweden\\
	E-mail: \email{pat@fysik.su.se}}
        
\author{Jan Conrad\\
	Oskar Klein Centre for Cosmoparticle Physics and\\
        Department of Physics, Stockholm University\\
        AlbaNova, SE-10691 Stockholm, Sweden\\
	E-mail: \email{conrad@fysik.su.se}}

\author{Joakim Edsj\"o\\
	Oskar Klein Centre for Cosmoparticle Physics and\\
        Department of Physics, Stockholm University\\
        AlbaNova, SE-10691 Stockholm, Sweden\\
	E-mail: \email{edsjo@fysik.su.se}}

\author{Lars Bergstr\"om\\
	Oskar Klein Centre for Cosmoparticle Physics and\\
        Department of Physics, Stockholm University\\
        AlbaNova, SE-10691 Stockholm, Sweden\\
	E-mail: \email{lbe@fysik.su.se}}

\author{Christian Farnier\\
	Laboratoire de Physique Thorique et Astroparticules, CNRS/IN2P3, Universit\'e Montpellier II, CC 70, Place Eugne Bataillon, F-34095 Montpellier Cedex 5, France\\
        E-mail: \email{farnier@in2p3.fr}}

\author{Yashar Akrami\\
	Oskar Klein Centre for Cosmoparticle Physics and\\
        Department of Physics, Stockholm University\\
        AlbaNova, SE-10691 Stockholm, Sweden\\
	E-mail: \email{yashar@fysik.su.se}}

\preprint{}	

\abstract{The dwarf galaxy Segue 1 is one of the most promising targets for the indirect detection of dark matter.  Here we examine what constraints 9 months of \FLAT gamma-ray observations of Segue 1 place upon the Constrained Minimal Supersymmetric Standard Model (CMSSM), with the lightest neutralino as the dark matter particle.  We use nested sampling to explore the CMSSM parameter space, simultaneously fitting other relevant constraints from accelerator bounds, the relic density, electroweak precision observables, the anomalous magnetic moment of the muon and $B$-physics.  We include spectral and spatial fits to the \Fermi observations, a full treatment of the instrumental response and its related uncertainty, and detailed background models.  We also perform an extrapolation to 5 years of observations, assuming no signal is observed from Segue 1 in that time.  Results marginally disfavour models with low neutralino masses and high annihilation cross-sections.  Virtually all of these models are however already disfavoured by existing experimental or relic density constraints.}


\begin{document}

\section{Introduction}

The identity of dark matter is one of the most compelling problems facing modern physics.  A wealth of viable theoretical candidates have been put forward (see e.g. \cite{Jungman96, Bergstrom00, Bertone05, Bergstrom09}), with the majority based on extensions to the standard model (SM) of particle physics.  One of the more durable suggestions is that dark matter consists of weakly-interacting massive particles (WIMPs), thermally produced in the early universe and therefore naturally present in approximately the right cosmological abundance.  Models of supersymmetry (SUSY) where $R$-parity is conserved provide a prototypical WIMP candidate in the lightest neutralino.  Low-energy SUSY is also highly attractive because it generically solves the SM hierarchy problem whilst simultaneously providing a favourable framework for gauge-coupling unification and electroweak symmetry breaking \cite{BaerTata, Aitchison}.

Because the neutralino is a Majorana particle, its self-annihilation opens a potential channel for discovery via the observation of annihilation products like photons, hadrons and leptons.  Self-annihilation rates are proportional to the square of the particle density, so any environment with a high density of dark matter is a good prospective target.  In practice, expected backgrounds from different targets strongly influence their suitability for such indirect detection.  Dwarf spheroidal galaxies have recently emerged as leading targets for gamma-ray detection of dark matter \cite{Strigari08,Bringmann09,Pieri09,Essig09,Lombardi09}, thanks to their high mass-to-light ratios \cite{Strigari08,Simon07,Belokurov09,Geha09} and small expected astrophysical backgrounds.  

Segue 1 is probably the most promising object in this respect \cite{Martinez09,Kuhlen09}, due to its extreme dark matter domination ($M/L\approx 1320$), relative proximity (23\,kpc) and high latitude \cite{Geha09}.  As with other dwarf galaxies, constraining the density profile of dark matter in Segue 1 is difficult; being small and faint, very few stars are available to act as kinematic tracers of the gravitational potential.  Its spatial superimposition upon the leading arm of the Sagittarius stream \cite{B07} complicates matters further, as do the partially degenerate impacts of dark matter, bulk rotation and magnetic fields upon the stellar velocity dispersion \cite{Xiang09}.  Indeed, the status of Segue 1 as a dwarf galaxy rather than a star cluster, and therefore its domination by dark matter, have been called into question \protect\cite{B07,Niederste09}.  We will assume here that it is indeed a galaxy, an assertion strongly supported by further recent (but as yet unpublished) spectroscopic data \protect\cite{GehaTeVPA}.  These new data should also significantly reduce the uncertainty associated with the density profile of dark matter within Segue 1.  As of the time of writing, the best available estimate of this profile comes from Markov-Chain Monte Carlo (MCMC) scans of halo parameters and corresponding solutions to the Jeans equation, based on line-of-sight velocities of 24 stars in Segue 1 \cite{Martinez09}.

The Large Area Telescope (LAT; \cite{Atwood09}), aboard the \Fermi satellite, is a high-energy, pair-conversion gamma-ray space telescope.  The LAT is designed to operate predominantly in survey mode, and has been doing so since August 4, 2008.  With its energy range (20\,MeV to over 300\,GeV) and high spatial and spectral resolution ($\Delta E / E \approx 12\%$, point-spread function $< 0.1\deg$ at 100\,GeV), the LAT is well-suited to gamma-ray searches for dark matter annihilation.  A major undertaking within the LAT collaboration has been to try to discover or place limits upon theories of dark matter using \Fermi observations of Milky Way dwarf galaxies and satellites, the Galactic centre, the Galactic halo and extragalactic sources \cite{DMNPprelaunch, Meurer09, DwarfRICAP, DwarfCatI, SatTeVPA}.  The detector design also facilitates direct observation of cosmic-ray electrons \cite{Fermielectron}, another possibly relevant channel for dark matter indirect detection.

By any measure, SUSY is an extensive and highly developed addition to the SM, giving rise to a wealth of potential experimental signatures beyond dark matter.  Any explanation of dark matter as a neutralino must therefore satisfy a host of other phenomenological constraints.  Even within the minimal supersymmetric extension of the SM (the MSSM), the low-energy phenomenology of the theory strongly depends upon the particular parameterisation employed in the soft SUSY-breaking sector, and the specific values of the chosen parameters.  Given that the most general soft SUSY-breaking Lagrangian in the MSSM has over a hundred free parameters, one must choose some reduced parameterisation in order to make any progress in fitting experimental data.  One approach is to employ a low-energy effective Lagrangian for the soft breaking terms, with various parameters set to zero or made equal for computational convenience (and in order to avoid experimental constraints on e.g. flavour-changing neutral currents).  The alternative is to choose a specific breaking scheme, such as gravity mediation in minimal supergravity (mSUGRA)\cite{Chamseddine82}, gauge mediation (GMSB)\cite{Giudice99} or anomaly mediation (AMSB)\cite{Randall99}, where a small number of breaking parameters are defined and unified at some high energy, and the masses and couplings are run down to low energy using the renormalisation group equations (RGEs) in order to obtain phenomenological predictions.

In this paper we focus on the Constrained MSSM (CMSSM) as a convenient example of one such high-energy parameterisation.  This scheme is defined at the gauge coupling unification scale ($\sim10^{16}$\,GeV) in terms of 4 free continuous parameters and one sign:
\begin{equation}
\{m_0, m_\frac{1}{2}, A_0, \tan\beta, \mathrm{sgn}\,\mu\}.
\end{equation}
Here $m_0$ is the universal scalar mass, $m_\frac{1}{2}$ the gaugino mass parameter, $A_0$ the trilinear coupling between Higgs bosons, squarks and sleptons, $\tan\beta$ the ratio of vacuum expectation values of up-type and down-type Higgs bosons, and $\mathrm{sgn}\,\mu$ the sign of the Higgs mixing parameter in the superpotential.  We choose $\mu$ to be positive throughout this paper.  The magnitude of $\mu$ is set by the requirement that SUSY breaking radiatively induces electroweak symmetry breaking; in this sense the CMSSM differs slightly from mSUGRA (where electroweak symmetry breaking is not strictly part of the definition and $\tan\beta$ is swapped for the parameter $B$ -- see e.g. \cite{BaerTata}), but for nearly all intents and purposes the two can be considered equivalent.

The CMSSM possesses a number of distinct regions where the relic density of the lightest neutralino matches the observed dark matter abundance (see e.g.~\cite{Martin,Allanach06} and references therein).  The majority of the CMSSM parameter space results in too high a relic density; regions producing the correct amount of dark matter are those where some channel of neutralino destruction is especially efficient.  Until recently, most analyses focused on the so-called bulk region at low $m_0$ and $m_\frac{1}{2}$, where neutralino annihilation proceeds efficiently by exchange of light sleptons.  This region is now mostly ruled out by collider limits on sparticle masses and difficulty in meeting Higgs mass limits when both $m_0$ and $m_\frac{1}{2}$ are small.  The stau coannihilation region occurs at low $m_0$, where the stau is almost degenerate in mass with the lightest neutralino.  Here the correct relic density is achieved via co-annihilations between the two sparticles rather than any increase in the neutralino self-annihilation cross-section.  A similar situation occurs in the stop coannihilation region, which exists at large negative $A_0$.  The stau coannihilation region is still viable, but stop coannihilation is disfavoured by low-energy experiments and Higgs constraints.  In the focus point region at large $m_0$, the lightest neutralino picks up a significant Higgsino component, opening new annihilation channels and boosting certain coannihilations.  Finally, small `funnels' of parameter space exist where neutralino annihilation can be increased by a mass resonance with one of the MSSM Higgs particles (i.e.~where the Higgs in question has roughly twice the mass of the lightest neutralino).  The focus point and funnel regions are still allowed by present experimental constraints.

Scanning of MSSM parameter spaces is nowadays a highly developed art.  Starting from simple grid and random scans \cite{Drees93, Baer96, Ellis01, Ellis04, Roszkowski01} within slices of the mSUGRA parameter space, efforts expanded to MCMC searches of the full CMSSM/mSUGRA space \cite{Baltz04}, later also including the most important SM uncertainties \cite{Ruiz06, Allanach06}.  As nested sampling \cite{Skilling04, MultiNest} has come to replace the MCMC as the scanning technique of choice, hope has risen that MSSM scans might now finally be globally convergent \cite{Trotta08, AbdusSalam09a}.  New results with genetic algorithms \cite{Akrami09}, however, suggest that current scanning techniques may yet have some distance to go in this respect.  Some authors have begun to focus on higher-dimensional low-energy effective MSSM parameterisations \cite{AbdusSalam09a, Cotta09, Berger09}, which provide for a broader range of phenomenological consequences but are almost impossible to scan effectively without sophisticated algorithms and substantial supercomputing resources.  Explorations of SUSY-breaking schemes beyond mSUGRA have also been carried out lately using similar parameter scans \cite{Roszkowski09b, AbdusSalam09b, Buchmueller08, Buchmueller09}, as have investigations of next-to-minimal SUGRA \cite{Balazs08, Balazs09, Lopez09}.  SUSY scans are generally either based on the Bayesian posterior probability \cite{Ruiz06, Trotta08, Roszkowski09b, AbdusSalam09a}, the direct use of the frequentist likelihood \cite{Ellis04, Allanach06, Buchmueller08} (usually by a $\chi^2$ analysis), or a simple `in-or-out' approach to individual points being permitted by experimental data \cite{Drees93, Baer96, Ellis01, Baltz04}.

MSSM scans have thus far focused on the constraints provided by particle experiments and the dark matter relic density determined from the microwave background, sometimes to produce corresponding predictions for astronomical observations (e.g. \cite{Martinez09, Roszkowski09a, Trotta09, SS09}).  To our knowledge, none have so far included actual constraints from searches for annihilating dark matter; this is no doubt because such constraints have only recently come within a reasonable distance of model predictions.

In this paper, we include the first 9 months of the search for dark matter annihilation in Segue 1 with \Fermi in explicit CMSSM parameter scans.  We use spectrally and spatially resolved photon counts observed by the LAT to directly assess the likelihood of the different regions in the CMSSM parameter space, then combine these with laboratory and cosmological data to perform global fits to the model parameters.  We also provide a predicted impact on the parameter space after 5 years of observations.  In Sect.~\ref{analysis} we describe our analysis techniques, before presenting results in Sect.~\ref{results} and conclusions in Sect.~\ref{conclusions}.

\section{Analysis}
\label{analysis}

\subsection{Gamma-rays from neutralino annihilation in dwarf galaxies}
\label{annihilation}

The expected differential gamma-ray flux per unit solid angle from a source of neutralino annihilations is (see e.g. \cite{Bergstrom98}) 
\begin{equation}
\label{fluxeqn}
\frac{\mathrm{d}\Phi}{\mathrm{d}E\mathrm{d}\Omega} = \frac{1 + BF}{8\pi m_\chi^2} \sum_f \frac{\mathrm{d}N^\gamma_f}{\mathrm{d}E}\sigma_f v \int_\mathrm{l.o.s.} \rho_\chi^2(l) \mathrm{d}l.
\end{equation}
Here $m_\chi$ is the neutralino mass, $BF$ is the boost factor due to any unresolved substructure in the source, $f$ labels different annihilation final states, $\mathrm{d}N^\gamma_f/\mathrm{d}E$ is the differential photon yield from any particular final state, $\sigma_f$ is the cross-section for annihilation into that state, $v$ is the relative velocity between neutralinos, and the integral runs over the line of sight to the source.  In the absence of any bound states (i.e. Sommerfeld enhancements), massive neutralinos move so slowly that they can effectively be considered to collide at rest, allowing $\sigma_f v$ to be replaced with the velocity-averaged term in the zero-velocity limit, $\langle\sigma_f v\rangle_0$.

Three main channels contribute to the spectrum of neutralino annihilation.  Through loop processes, annihilation can proceed directly into two photons \cite{Bergstrom88,Bergstrom97}
\begin{equation}
\frac{\mathrm{d}N^\gamma_{\gamma\gamma}}{\mathrm{d}E} = 2\delta(E - m_\chi),
\end{equation}
or into a Z boson and a photon \cite{Ullio98}
\begin{equation}
\frac{\mathrm{d}N^\gamma_{Z\gamma}}{\mathrm{d}E} = \delta(E - m_\chi + \frac{m_Z^2}{4m_\chi}),
\end{equation}
giving a monochromatic gamma-ray line.  A hard spectrum can also be produced by the so-called internal bremsstrahlung (consisting of final-state radiation and virtual internal bremsstrahlung), generated when a photon is emitted from a virtual particle participating in the annihilation diagram \cite{Bringman08}.  Finally, continuum gamma-rays can be produced by annihilation into quarks, leptons and heavy gauge bosons (including the $Z$ from the $Z\gamma$ line), which subsequently decay via $\pi^0$ to softer photons.  The cross-sections and resultant spectral yields for each of these processes are directly calculable from the SUSY parameters which define a point in e.g. the CMSSM parameter space (after appropriate RGE running).  We use \textsf{DarkSUSY} \cite{darksusy} for this calculation. 

The integral and boost factor in Eq.~\ref{fluxeqn} are determined by the dark matter distribution in the astrophysical source.  We use the Einasto profile \cite{nfwsmooth}
\begin{equation}
\label{einasto}
\rho(r) = \rho_\mathrm{s}\exp\left\{-2n\left[\left(\frac{r}{r_\mathrm{s}}\right)^\frac{1}{n} -1 \right]\right\}
\end{equation}
to describe the average dark matter content of Segue 1, where $n$ is the Einasto index and $r_\mathrm{s}$ and $\rho_\mathrm{s}$ are the scale radius and density, respectively.  This profile is somewhat more conservative than the traditional NFW \cite{NFW}, in the sense that it is less steep in the central regions, leading to generally better agreement with observations of various dark matter halos \cite{nfwsmooth,Merritt06,Gao08}.  It is also slightly more dense at intermediate radii.  The adopted form of the density profile actually makes little overall difference to the expected flux.  This is because in general, dwarf galaxies will appear either as point sources or very close to pointlike to the LAT, meaning that observations mostly probe the full halo rather than just the central cusp.

We use the best-fit values of the scale radius and scale density found by Martinez et al.~\cite{Martinez09} in their recent fits to stellar kinematic data ($r_\mathrm{s} = 0.07$\,kpc, $\rho_\mathrm{s}=3.8$\, GeV\,cm$^{-3}$).  Since Martinez et al.~found no preference for a particular Einasto index, we adopt the central value considered in their scans, $n=3.3$.  We note that the fits were not only influenced by the kinematic data, but also by a theoretical prior imposed by assuming the same correlation between $r_\mathrm{s}$ and $\rho_\mathrm{s}$ as seen in subhalo populations of theoretical $N$-body simulations of cold dark matter structure formation.  Whilst this presents no real problem, it is encouraging to see that additional kinematic data \cite{GehaTeVPA} largely dominate the prior in more recent fits.  The same authors performed an extensive investigation of the possible substructure boosts in Segue 1, showing that all $BF$ values between 0 and $\sim$70 are compatible with kinematic data and small-scale structure predictions within the CMSSM, with the most likely value depending strongly on the particular model employed for the concentration-mass relation.  We therefore employ two indicative values for the boost factor: a rather pessimistic case, $BF=1$, and an optimistic case, $BF=50$.  It is important to note that $BF=0$ has a very low probability in the results of Martinez et al. 

\subsection{Observations and instrumental considerations}
\label{observations}

We considered photon events observed in a 10 square degree, stereographically-projected section of the sky centred on Segue 1 ($\mathrm{RA, Dec.} = 151.763\deg, 16.074\deg$ \cite{Geha09}).  We applied cuts on event zenith angles ($\theta<105\deg$), energies ($100\,\mathrm{MeV}<E<300\,\mathrm{GeV}$) and identifications (only `diffuse class' events -- see \cite{Atwood09}).  All data were processed using the same reconstruction algorithms and instrument response functions (IRFs) as the publicly-released first-year data.  Counts and corresponding exposures were placed into $64 \times 64$ spatial and 14 logarithmic energy bins.  The resultant energy-integrated map of photon counts is shown in Fig.~\ref{fig1}.

The \FLAT IRFs consist of the effective area, point-spread function (PSF) and energy dispersion.  We factored the effective area of the telescope into our calculations of the exposure for each bin of observed photon counts, using the the standard analysis tools available from HEASARC, specifically \textsf{ScienceTools 9.11}\footnote{Available at \href{http://fermi.gsfc.nasa.gov/ssc/data/analysis/software/}{\tt http://fermi.gsfc.nasa.gov/ssc/data/analysis/software/}.}.  We convolved our modelled gamma-ray fluxes with the PSF and energy dispersion of the LAT using the publicly-available Fortran90 library \textsf{FLATlib} \cite{FLATlib}, which was designed specifically for performing this task quickly enough to be useful in MSSM scans.  Full \FLAT IRFs are defined not only as a function of photon energy, but impact angle with respect to the telescope zenith (and even azimuthal angle, though the dependence is weak).  \textsf{FLATlib} achieves its fast convolution by averaging the IRFs over impact angles, allowing the integral over the PSF to be cast as a true convolution and performed by fast spectral methods.  The energy integral cannot be performed in a similar way, because all three IRFs remain energy-dependent.  \textsf{FLATlib} performs this integral explicitly, using a fast importance-sampling technique which utilises the rough resemblance of the energy dispersion function to a Gaussian.  For the sake of computational speed, we truncated the PSF at a width of $3.2\deg$ in the scans we present in this paper.  With the present IRF set, this is well beyond the LAT's 95\% containment resolution at e.g. 100\,GeV ($\approx 0.3\deg$) or even 1\,GeV ($\approx 2\deg$).  

\FIGURE{
  \centering
  \includegraphics[width=70mm, trim = 50 90 50 200, clip=true]{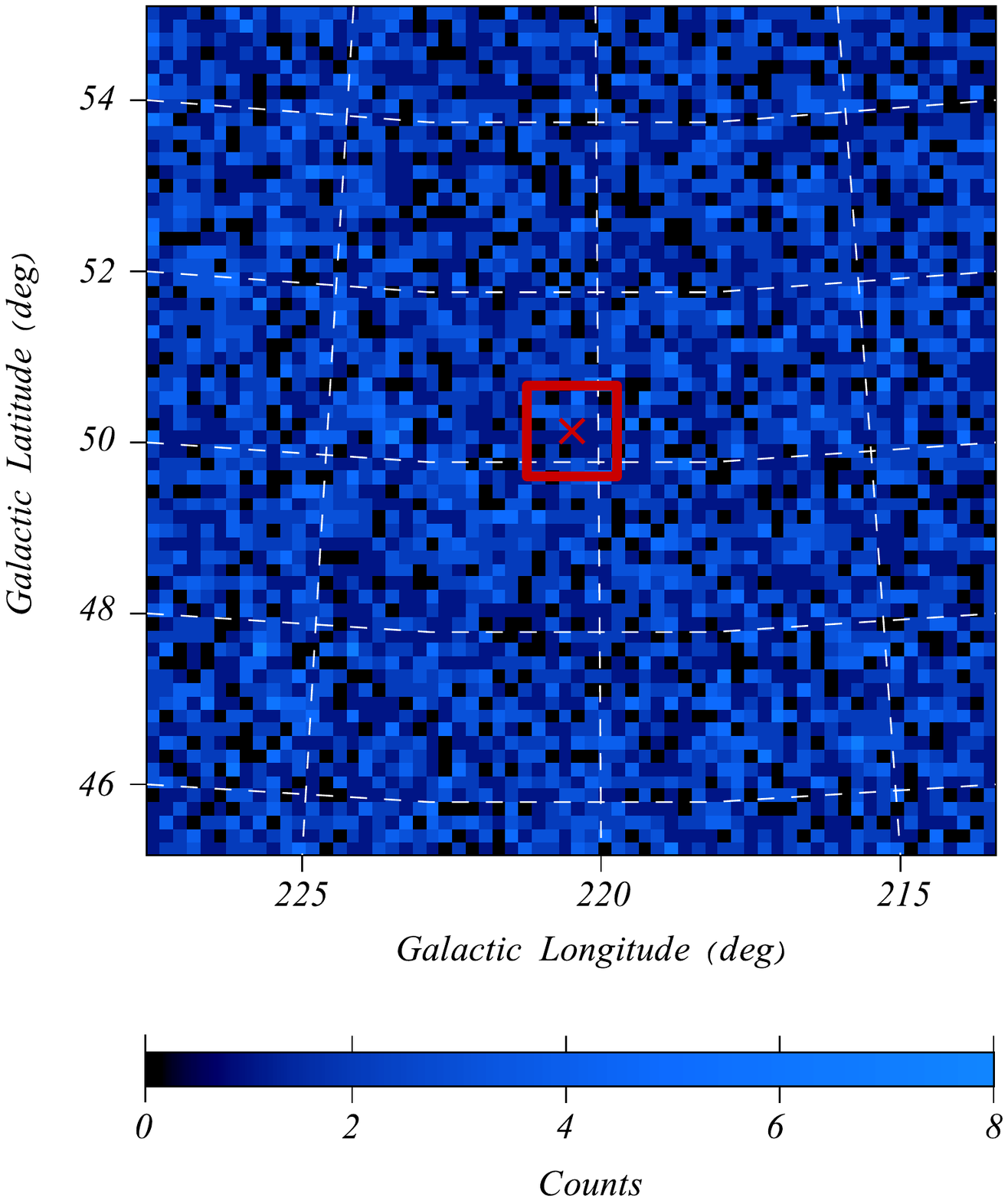}
  \caption{Photon counts observed by \Fermi in the region around Segue 1 during the first 9 months of LAT operation in all-sky survey mode.  Counts are integrated over all energies between 100\,MeV and 300\,GeV.  The red cross shows the exact location of the centre of Segue 1, and the red box shows the region included in our likelihood calculations.}
  \label{fig1}
}

\subsection{Likelihoods from Segue 1}
\label{likelihoods}

The expected spatial extent of Segue 1 in the gamma-ray sky, if it shines with dark matter annihilation, is comparable to the width of the LAT PSF.  This puts \mbox{Segue 1} on the borderline between a predicted point source and a predicted extended source.  For every set of CMSSM parameters, we computed model spectra at each pixel in the inner $6\times 6$ square shown in red in Fig.~\ref{fig1}.  We took care to explicitly integrate the density profile over the innermost $2\times 2$ pixels as a whole, so as to correctly capture the contribution of the very centre of the galaxy (located at their vertex).  We compared the predicted spectra with the observed ones in each of the 36 pixels to obtain a likelihood based on 504 data points, which we then included in the total likelihood for that point in our CMSSM scan.  We chose only to include the inner 36 pixels in the CMSSM likelihood simply because these are the only pixels where there is a predicted signal at any significant level.

All modelled spectra explicitly included contributions from gamma-ray lines, internal bremsstrahlung and continuum radiation.  To properly model the observed event counts in the region around Segue 1, we also took the Galactic and isotropic diffuse emissions into account.  We used a preliminary form of the \textsf{GALPROP} fit to the emission observed by \Fermi \cite{DiffuseCatI} to describe the Galactic diffuse emission.  The contribution of the isotropic diffuse emission, presumably originating from extragalactic sources, is much weaker and depends on the Galactic diffuse model adopted.  To describe this, we adopted an isotropic power law model with index $-2.1$, derived from EGRET observations by Sreekumar et al.~\cite{Sreekumar98}.  The models recommended by the LAT team were updated recently, and released to the Fermi Science Support Centre.  At the $\sim$50$^\circ$ latitude of Segue 1, the differences between the old and new models are not important for this analysis.  The normalisations for both backgrounds were set to the best-fit values obtained in the preliminary 9-month LAT dwarf upper-limit analysis \cite{DwarfRICAP, DwarfTeVPA}, based on the full 10 square degree region of interest rather than just the inner 36 pixels included in our likelihoods.  No sources were detected in this region in the first 9 months of LAT operation.

Because of the very low statistics observed in LAT photon counts towards Segue 1, a $\chi^2$ estimation of the likelihood is inappropriate in this case.  We calculated the likelihood using a binned Poissonian measure
\begin{equation}
\mathcal{L} = \prod_j \frac{\theta_j^{n_j}\mathrm{e}^{-\theta_j}}{n_j!},
\end{equation}
or, recast in the more familiar minus log-likelihood form (analogous to half the $\chi^2$),
\begin{equation}
-\ln\mathcal{L} = \sum_j \left[\theta_j + \ln(n_j!) - n_j\ln(\theta_j)\right].
\end{equation}
Here $n_j$ and $\theta_j$ are the observed and predicted number of counts respectively, in the $j$th bin.  This prescription clearly accounts for statistical errors by definition, but including systematic errors is less obvious.  To do so, one can marginalise over an assumed probability density function (PDF) of a systematic error in a semi Bayesian manner, treating it as a nuisance parameter.  If we consider a systematic error that has the impact of consistently rescaling the observed number of counts as $n_j\to \epsilon n_j$ (i.e. a constant percentage systematic error $|1-\epsilon|$), and assume a Gaussian form with width $\sigma_\epsilon$ for the PDF of $\epsilon$, the marginalised log-likelihood is (see e.g. \cite{Conrad03})
\begin{align}
-\ln\mathcal{L} & =  - \sum_j \ln\left\{\frac{1}{\sqrt{2\pi}\sigma_\epsilon}\int_0^\infty \frac{(\epsilon\theta_j)^{n_j}\mathrm{e}^{-\epsilon\theta_j}\exp\left[-\frac{1}{2}\left(\frac{1-\epsilon}{\sigma_\epsilon}\right)^2\right]}{n_j!}\mathrm{d}\epsilon\right\} \\
& = - \sum_j \ln\left\{\frac{\theta^{n_j}}{\sqrt{2\pi}\sigma_\epsilon n_j!}\int_0^\infty \epsilon^{n_j}\exp\left[-\epsilon\theta_j-\frac12\left(\frac{1-\epsilon}{\sigma_\epsilon}\right)^2\right]\mathrm{d}\epsilon\right\}.
\end{align}
The integral is only analytically soluble for $\theta_j<\sigma_\epsilon^{-2}$, which is not generally true when dealing with small statistics; we performed it numerically for each likelihood evaluation.

We included estimated systematic errors from the LAT effective area ($f$) and our modelled spectra ($\tau$) by combining them in quadrature, i.e. $\sigma_\epsilon(E_j) = \sqrt{f(E_j)^2 + \tau^2}$.  Note the explicit energy dependence of $f$; for the present IRF set, $f(E_j)$ ranges from 10\% at 100\,MeV, to 5\% at 562\,MeV, to 20\% at 10\,GeV.  We interpolated between these values linearly, and assumed the edge values outside this range.  We tuned the importance sampling algorithm used by \textsf{FLATlib} using slower, more accurate standard numerical integration schemes, choosing a sampling efficiency for our specific problem that would introduce an overall systematic theoretical error $\tau$ of no more than 5\% in the normalisation of flux predictions.  Other systematic errors are no doubt also present in the theoretical predictions, but we expect the term from the fast integration to dominate.

\subsubsection{Extrapolation to 5 years of observations}
\label{methods5yr}

To make predictions about the impact of 5 years of LAT observations, we explicitly assume that no excess events will have been observed after this time.  There is no correct way to rescale Poissonian counts to longer timescales, so the Poissonian likelihood above cannot be used when extrapolating to longer observing times.  We instead set the `observed' number of photons equal to the number predicted by the background model, using rescaled 9-month exposures.  This prescription also avoids the erroneous shifts which confidence intervals based on Poissonian statistics can sometimes experience due to a downward statistical fluctuation of the background.  In this case the observed counts become a continuous instead of a discrete variable, so the problem of small statistics disappears.  The appropriate likelihood measure is then once more the $\chi^2$
\begin{equation}
\chi^2 = \sum_j\frac{\left(\Phi_{\mathrm{model},j} - \Phi_{\mathrm{observed},j}\right)^2}{\sigma_j^2} = \sum_j\frac{\left(\frac{\theta_j - n_j}{\mathcal{E}_j}\right)^2}{\sigma_{\mathrm{model},j}^2 + \sigma_{\mathrm{observed},j}^2},
\end{equation}
where $\Phi_{\mathrm{model},j}$ and $\Phi_{\mathrm{observed},j}$ are the predicted and observed fluxes, $\sigma_{\mathrm{model},j}$ and $\sigma_{\mathrm{observed},j}$ are their standard deviations, and $\mathcal{E}_j$ is the exposure.  The exposure is itself the product of the effective area and observing time.  The standard deviation of the predicted flux can be estimated as simply the product of the predicted flux and the percentage systematic theoretical uncertainty $\tau$ (5\% in our case -- see above), $\sigma_{\mathrm{model},j} = \tau\Phi_{\mathrm{model},j}$.  The standard deviation in the observed flux can be estimated from the standard deviation of the observed counts $\sigma_{n_j}$, and the uncertainty on the exposure $\sigma_{\mathcal{E}_j}$, giving
\begin{equation}
\sigma_{\mathrm{observed},j}^2 = \left(\frac{n_j}{\mathcal{E}_j}\right)^2\left(\frac{\sigma_{n_j}^2}{n_j^2} + \frac{\sigma_{\mathcal{E}_j}^2}{\mathcal{E}_j^2}\right).
\end{equation}
Since the underlying physical process is still Poissonian, the best estimate of $\sigma_{n_j}$ is in fact $\sigma_{n_j} = \sqrt{\theta_j}$.  Furthermore, since the uncertainty in the observing time is negligible, $\sigma_{\mathcal{E}_j}$ can be estimated as simply the percentage systematic error of the effective area $f(E_j)$ times the actual exposure, $\sigma_{\mathcal{E}_j} = f(E_j)\mathcal{E}_j$.  We then have
\begin{align}
\sigma_{\mathrm{observed},j}^2 & =  \left(\frac{n_j}{\mathcal{E}_j}\right)^2\left(\frac{\theta_j}{n_j^2} + f(E_j)^2\right)\\
& = \frac{\Phi_{\mathrm{model},j}}{\mathcal{E}_j} + \Phi_{\mathrm{observed},j}^2 f(E_j)^2,
\end{align}
giving
\begin{equation}
\chi^2 = \sum_j\frac{\left(\Phi_{\mathrm{model},j} - \Phi_{\mathrm{observed},j}\right)^2}{\frac{\Phi_{\mathrm{model},j}}{\mathcal{E}_j} + \Phi_{\mathrm{observed},j}^2 f(E_j)^2 + \tau^2\Phi_{\mathrm{model},j}^2}.
\end{equation}

We hasten to point out that constraints based on this extrapolation are probably overly conservative, as we assume the same background rejection, systematic errors and background model for both the 9-month analysis and the 5-year extrapolation.  Our overall understanding of the instrument will improve over time, as will our understanding of the background as \Fermi accumulates better statistics on the Galactic diffuse and extragalactic components, leading to correspondingly better constraints on the annihilation cross-section.  Kinematic constraints upon the dark matter density profile of Segue 1 will also improve in time \cite{GehaTeVPA, DwarfCatI}, which may impact constraints on CMSSM parameters.

\subsection{CMSSM scans}
\label{scans}

We scanned the CMSSM parameter space using a modified version of \textsf{\mbox{SuperBayeS} 1.35} \cite{Trotta08}, employing the \textsf{MultiNest} \cite{MultiNest} nested sampling algorithm with 4000 live points.  In the plots we show, all parameters except those shown on figure axes have been marginalised over in some way.  In the case of the frequentist profile likelihood, this is simply a matter of maximising the likelihood in the other dimensions of the parameter space.  In the case of the Bayesian posterior, the total posterior (prior times likelihood) is integrated over the other dimensions of the space (for a review see e.g.~\cite{TrottaBayesSky}).  Because we are somewhat more interested in the prior-independent profile likelihood than the marginalised posterior\footnote{We make the point, however, that both should be considered if one wants to gain as complete a picture as possible of the preferred regions in an insufficiently-constrained parameter space like the CMSSM.}, we prefer linear priors on the CMSSM parameters because they are flat relative to the likelihood, causing the sampling algorithm to proceed strictly according to the frequentist likelihood function.  The effects of alternative priors have already been discussed in detail for previous CMSSM scans \cite{Trotta08}.

We used \textsf{DarkSUSY 5.04} for the relic density and indirect detection computations.  This allowed us to calculate internal bremsstrahlung spectra, and improved the continuum spectrum and relic density calculations.  We also improved the interface between \textsf{SuperBayeS} and \textsf{DarkSUSY}, most notably pertaining to the energies at which some particle masses were defined.

Apart from the \Fermi data, the experimental data and nuisance parameters which we included in scans were identical to those in \cite{Trotta08} and \cite{Akrami09}.  SM nuisance parameters were the top and bottom quark masses and strong and electromagnetic coupling constants.  Experimental data were precision electroweak measurements of SM parameters from the Large Electron-Positron collider (LEP), the relic density from 5-year Wilkinson Microwave Anisotropy Probe (WMAP) fits ($\Omega_\mathrm{DM} h^2 = 0.1099\pm 0.0062$ \cite{wmap5_dunkley}), LEP constraints on sparticle masses, LEP constraints on the Higgs mass, the anomalous magnetic moment of the muon ($g-2$), the $\bar{B}_s - B_s$ mass difference, and branching fractions of rare processes $b\to s\gamma$, $\bar{B}_u \to \nu\tau^-$ and $\bar{B}_s \to \mu^+\mu^-$.  Details can be found in \cite{Trotta08}.

In our chosen configuration, completing the integration over the LAT IRFs for a given point in the CMSSM parameter space required a similar order of magnitude in processing time as a relic density calculation.  Since the relic density computation is the main bottleneck in MSSM scans, this meant that scans took roughly twice as much total processor time to complete as a standard \textsf{SuperBayeS} run.  One advantage of \textsf{FLATlib}, however, is that it can employ the multi-threaded version of the FFTW library \cite{FFTW}, allowing the IRF integration to be performed with a considerably greater degree of parallelisation than the present relic density routines.

\section{Results and Discussion}
\label{results}

\subsection{Fits to \Fermi data only}
In Fig.~\ref{fig2} we show results of scans where the likelihood function only included \Fermi data, LEP measurements of nuisance parameters and the requirements of physicality (the absence of tachyons, that the neutralino is the lightest SUSY particle, and that electroweak symmetry breaking is induced by SUSY breaking).  Preferred values of the neutralino self-annihilation cross-section and mass are shown for scans including 9 months of data, scans including the extrapolation to 5 years of data, and a control case without any \Fermi data.  Preferred regions are also given for both the pessimistic and optimistic boost factors discussed in Sect.~\ref{annihilation}.

One apparent feature of Fig.~\ref{fig2} is the lack of viable models with large annihilation cross sections for large neutralino masses.  This feature is present simply because the annihilation cross section goes as $m_{\chi}^{-2}$, causing it to fall off at higher masses.

Given the absence of any observed signal from Segue 1, \Fermi data clearly disfavours models with the highest cross-sections and lowest masses.  This is expected, since higher cross-sections and lower masses lead to a larger predicted signal.  That constraints are best at lower neutralino masses is also consistent with the falling sensitivity of the LAT with energy above about 50\,GeV, and the reduced source statistics at higher energies.  The improvement in constraints when moving from the current 9 months of data to the 5-year predictions is also roughly what would be expected from a $\sqrt{t}$ improvement in sensitivity.  This shows that the two different likelihood estimators we employ give consistent results (we also checked this explicitly for 9 months of data, finding very good agreement).

Predictably, the adopted boost factor plays a large role in determining the extent of constraints brought to bear on the CMSSM by Segue 1.  In the most pessimistic scenario, 9 months of LAT observations have no impact on confidence regions, as all disfavoured cross-sections are larger than allowed by physicality arguments.  In the most optimistic scenario, the data disfavours all models with cross-sections greater than $\sim3\times10^{-25}$\,cm$^{3}$\,s$^{-1}$.  Improvements in constraints when moving from $BF=1$ to $BF=50$ are consistent with the factor of $51/2$ improvement in sensitivity expected from Eq.~\ref{fluxeqn}, as the most pessimistic constraints lie above the extent of contours in the upper middle panel of Fig.~\ref{fig2}.  Extrapolating to 5 years of observations, all values above $10^{-25}$\,cm$^3$\,s$^{-1}$ would be disfavoured, as would a region extending down below $10^{-26}$\,cm$^3$\,s$^{-1}$ at the lowest masses.  Once again, we caution the reader that this extrapolation does not take into account systematic improvements in the background and dark matter profile modelling after 5 years, nor in the LAT reconstruction algorithms (see Sect.~\ref{methods5yr}).

For comparison, in Fig.~\ref{fig2} we also show the previously-presented, preliminary 95\% confidence level upper limit from 9 months of LAT observations \cite{DwarfTeVPA}.  This limit was derived assuming annihilation proceeds only into $b\bar{b}$ pairs.  Apart from the obvious difference in overall strategy (upper limits from an assumed final state versus inclusion in explicit model scans), our analysis differs from the upper limit one in a number of ways.  The upper limit was derived assuming a point source for Segue 1, whereas we perform spatial fits; the upper limit is based upon an NFW rather than Einasto density profile, and does not include systematic errors nor a treatment of the energy dispersion.  

Nonetheless, the areas disfavoured in our scans are broadly consistent with the 9 month upper limit, a positive comment on the reliability of both analyses.  Our corresponding exclusions do however occur at somewhat higher cross-sections than in the upper limit analysis (i.e.~our exclusion region is above both the extent of coloured contours and the black line in the upper middle panel of Fig.~\ref{fig2}).  This is to be expected, as our ability to exclude models is degraded relative to the upper limit analysis by properly accounting for the systematic error in the effective area.  Because this error is energy-dependent, our exclusions also have a slightly different energy-dependence than the 95\% upper limit.

\FIGURE[t]{
  \begin{minipage}{0.31\columnwidth}
  \centering
  \includegraphics[width=\linewidth, trim = 20 170 0 200, clip=true]{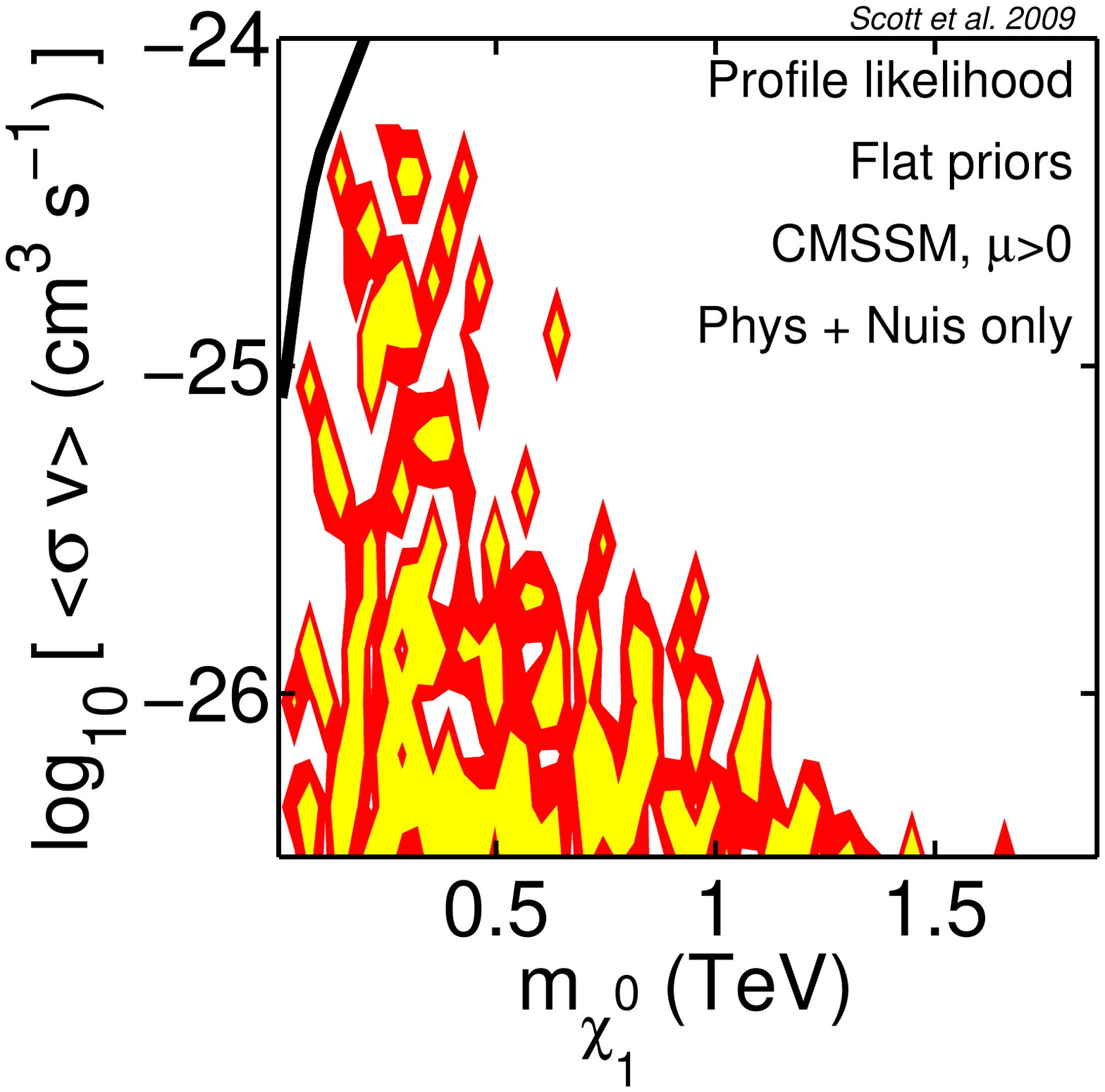}
  \end{minipage}
  \begin{minipage}{0.31\columnwidth}
  \includegraphics[width=\linewidth, trim = 20 170 0 200, clip=true]{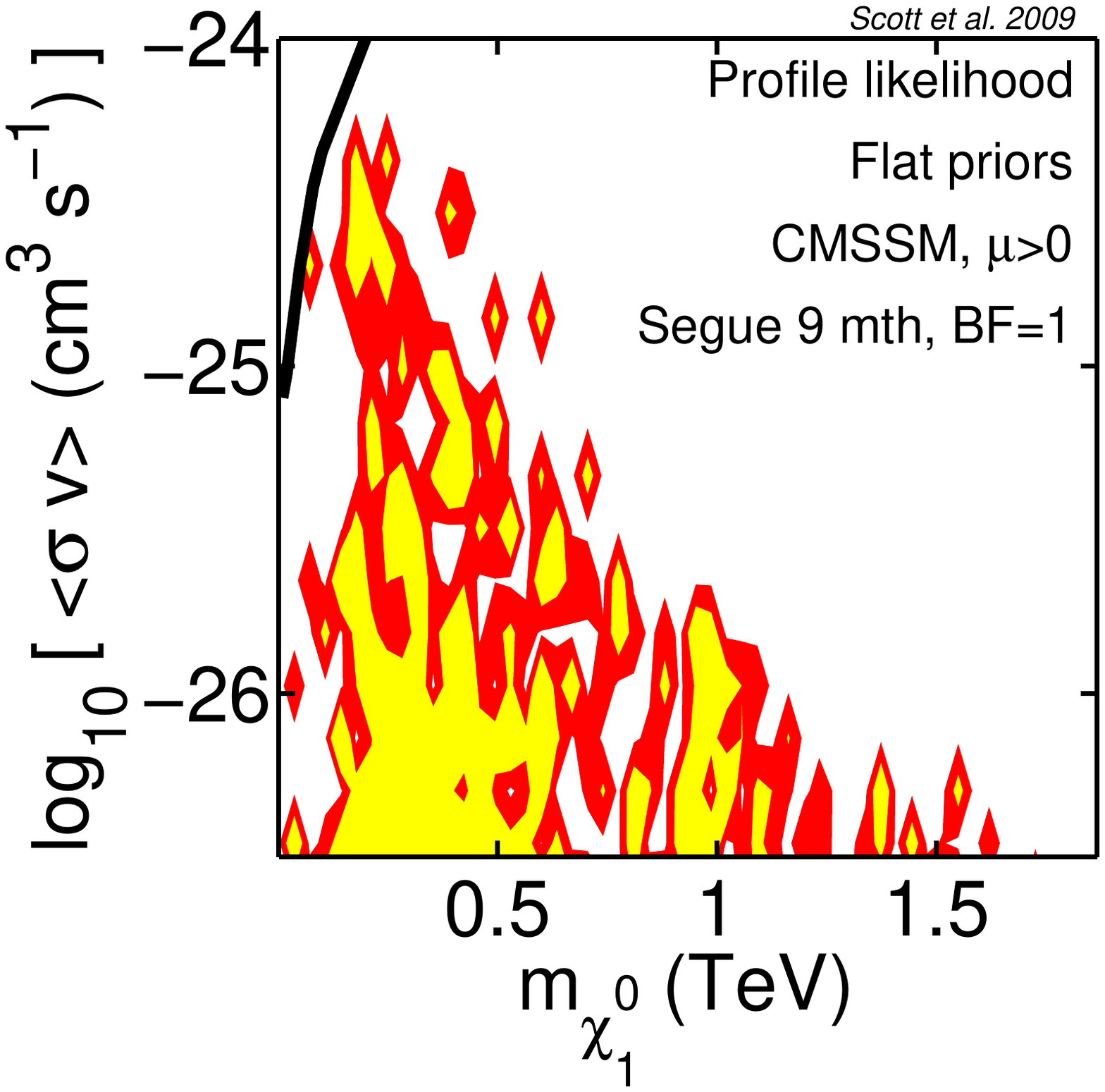}
  \includegraphics[width=\linewidth, trim = 20 170 0 200, clip=true]{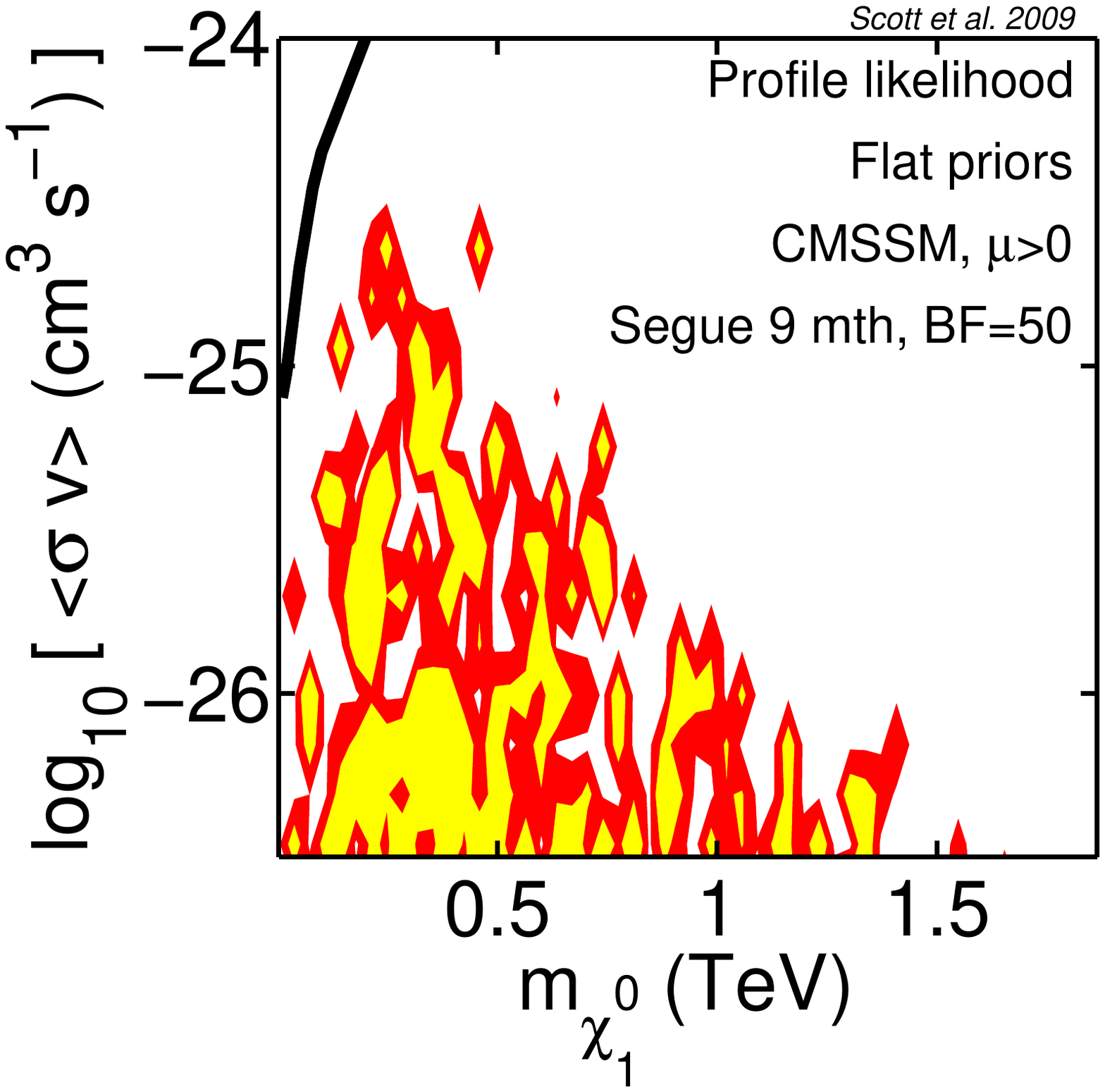}
  \end{minipage}
  \begin{minipage}{0.31\columnwidth}
  \includegraphics[width=\linewidth, trim = 20 170 0 200, clip=true]{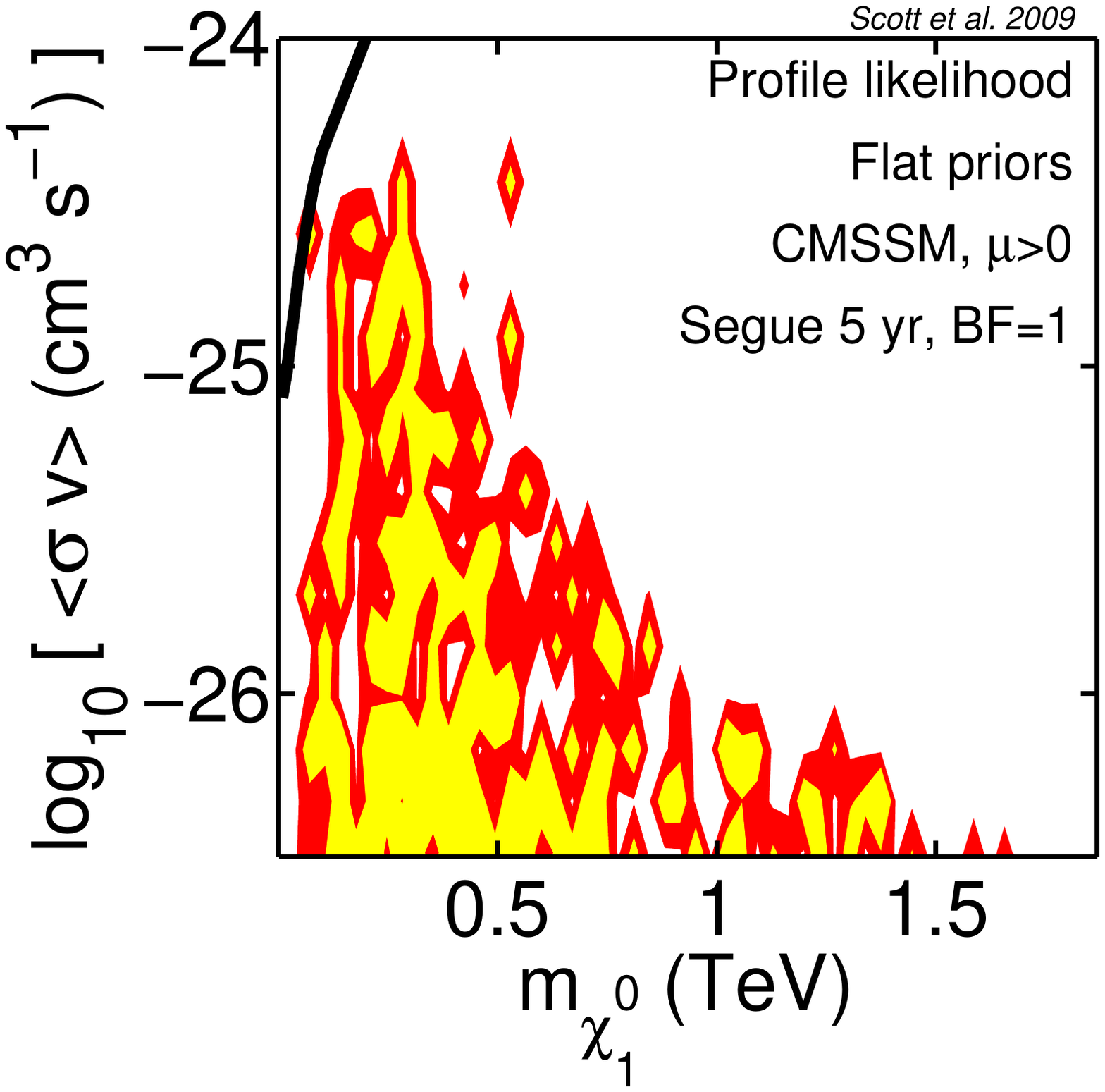}
  \includegraphics[width=\linewidth, trim = 20 170 0 200, clip=true]{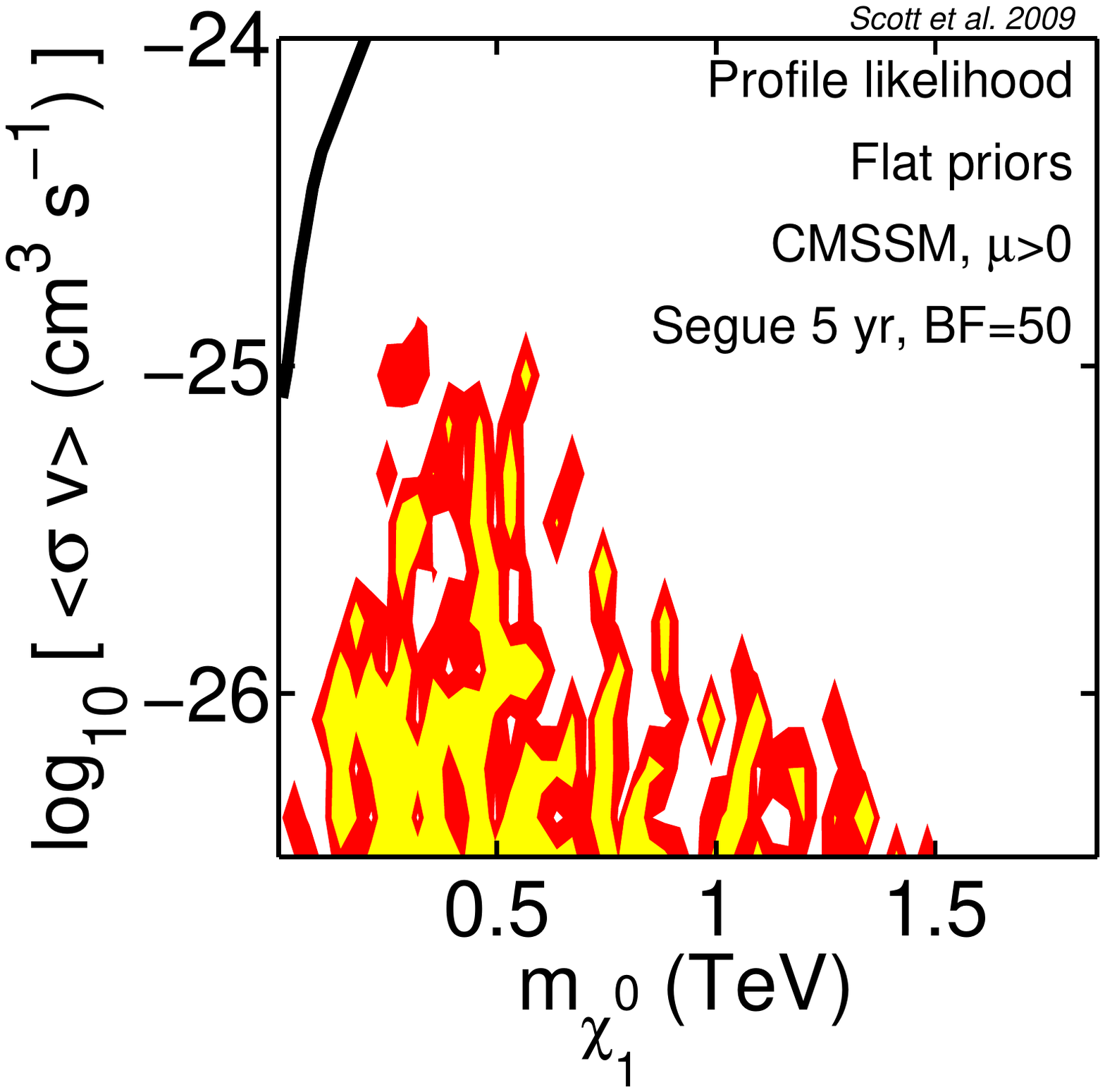}
  \end{minipage}
  \caption{Neutralino self-annihilation cross-sections in the CMSSM, in the zero-velocity limit.  \emph{Left}: with no constraining experimental data except measurements of SM nuisance parameters and physicality requirements.  \emph{Middle}: constraints provided by 9 months of \Fermi data on Segue 1, under the most pessimistic (\emph{top}) and optimistic (\emph{bottom}) assumptions about the substructure boost factor.  \emph{Right}: projected constraints after 5 years of \Fermi observations.  Colours indicate 68\% (yellow) and 95\% (red) confidence regions.  The preliminary 95\% confidence level upper limit on the annihilation cross-section from 9 months of \Fermi data, assuming 100\% WIMP annihilation into $b\bar{b}$ \cite{DwarfTeVPA}, is given for comparison (black curve).}
  \label{fig2}
}

It should be noted that the degree of substructure apparent in the confidence regions of Fig.~\ref{fig2} is unlikely to be physical, and is indeed probably something of an artefact of the scanning technique (i.e. `scanning noise').  In the absence of any constraint on the annihilation cross-section from the relic density, the vast majority of points providing a good fit to the included data lie at much lower cross-sections.  This prompts the scanning algorithm to concentrate its efforts there, leaving the region in which we are most interested somewhat poorly sampled.  From a Bayesian point of view, one would say that when the relic density is not included, this region sits well above the most likely annihilation cross-sections in the CMSSM, so is not meant to be very well sampled by the nested sampling technique.

Because only a small number of models are disfavoured by including just Segue data in the likelihood function, there is little overall impact on the favoured values of $m_0$, $m_\frac{1}{2}$, $A_0$ and $\tan\beta$ beyond what is allowed purely on physicality grounds.  We will show confidence regions from global fits only for these parameters.

\FIGURE[t]{
  \begin{minipage}{0.31\columnwidth}
  \centering
  \includegraphics[width=\linewidth, trim = 20 170 0 200, clip=true]{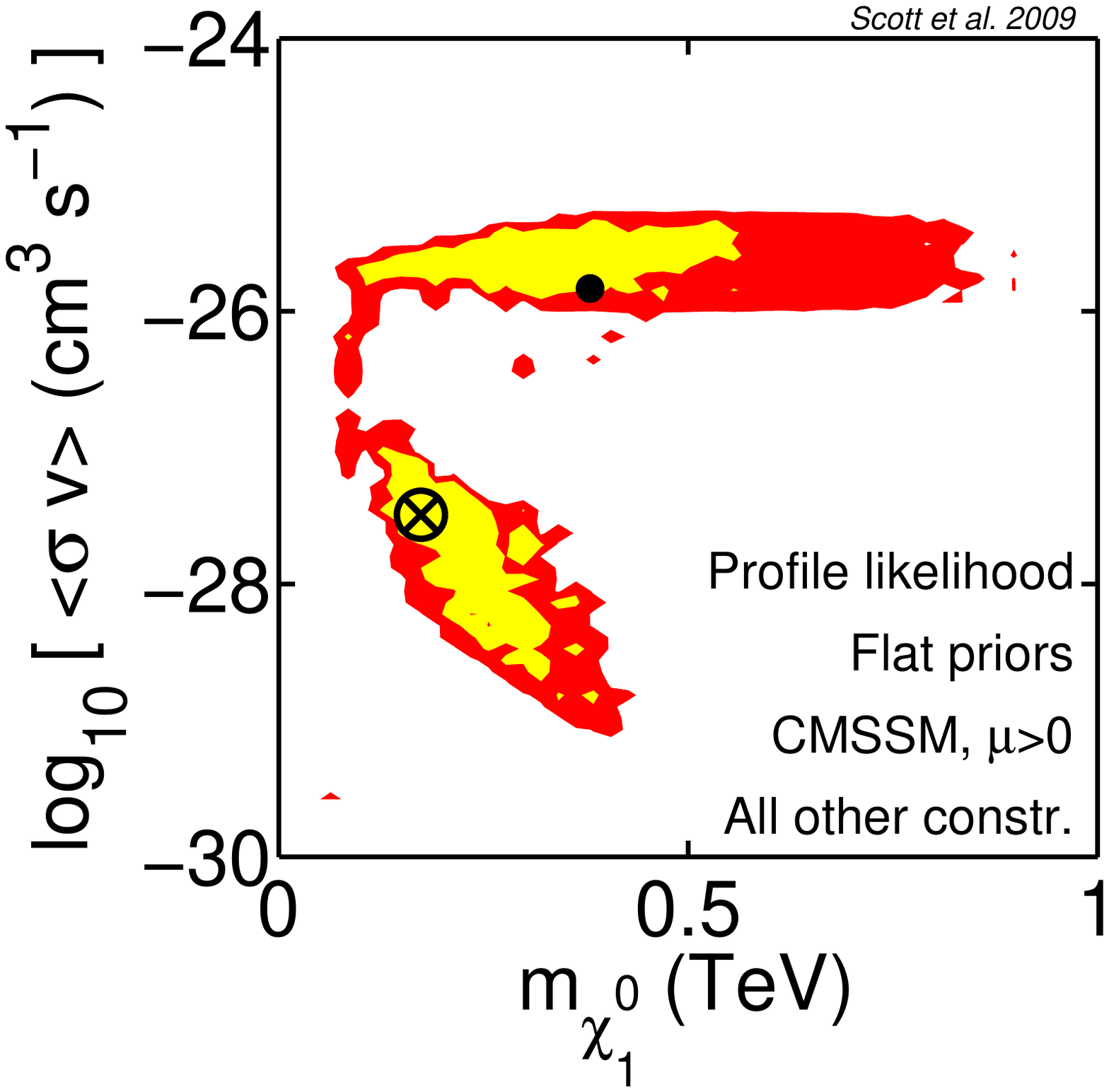}
  \includegraphics[width=\linewidth, trim = 20 170 0 200, clip=true]{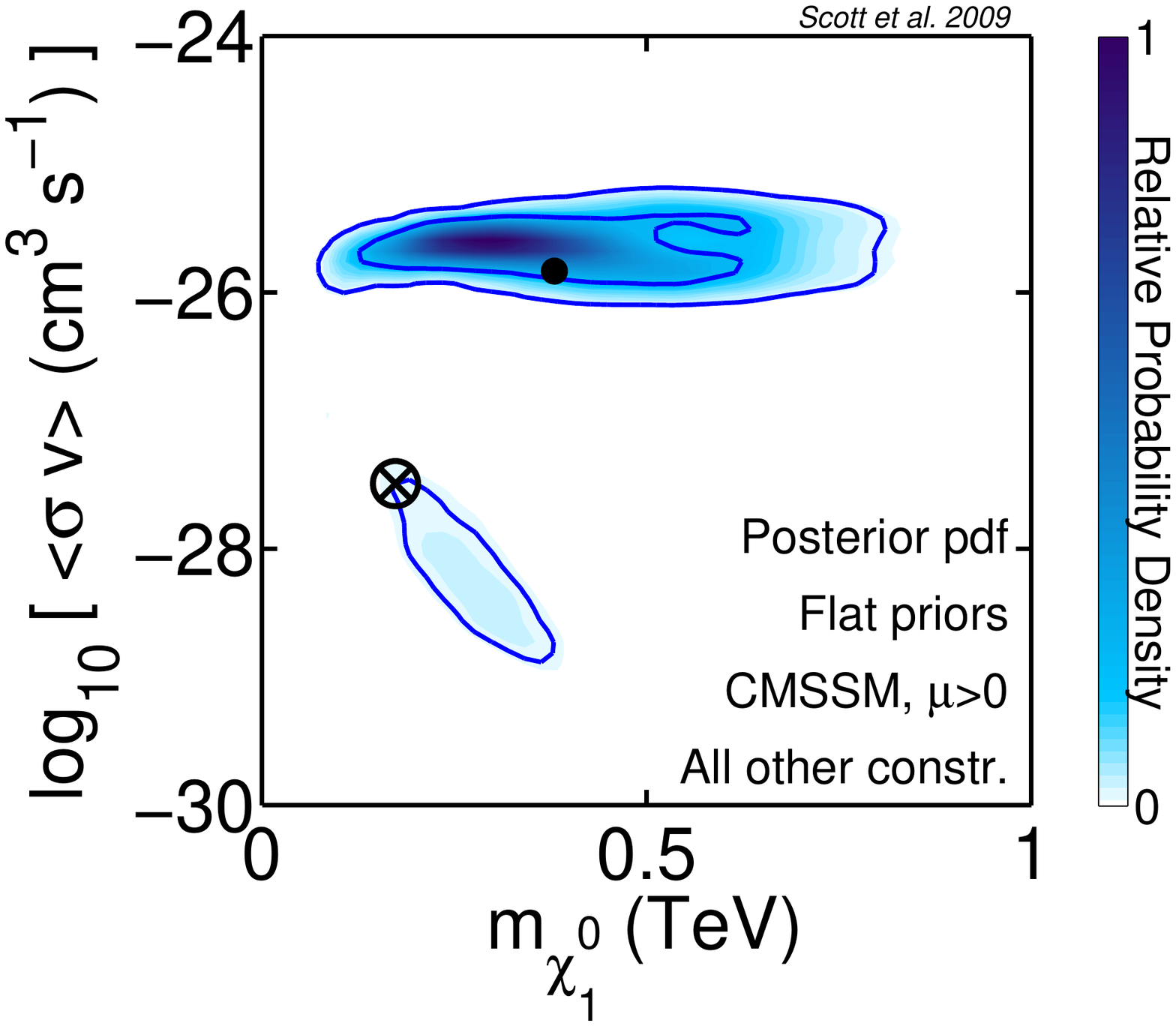}
  \end{minipage}
  \begin{minipage}{0.31\columnwidth}
  \includegraphics[width=\linewidth, trim = 20 170 0 200, clip=true]{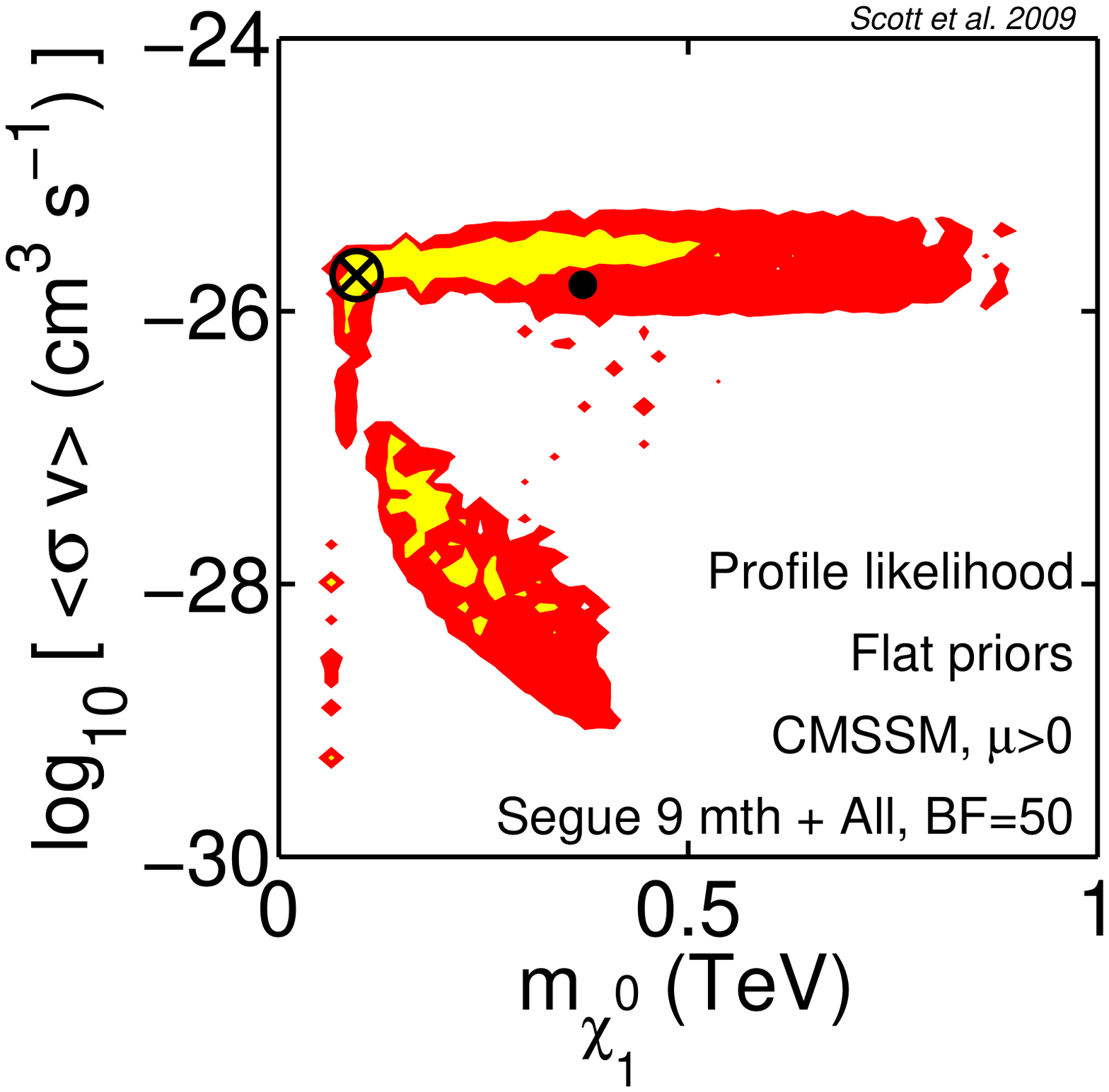}
  \includegraphics[width=\linewidth, trim = 20 170 0 200, clip=true]{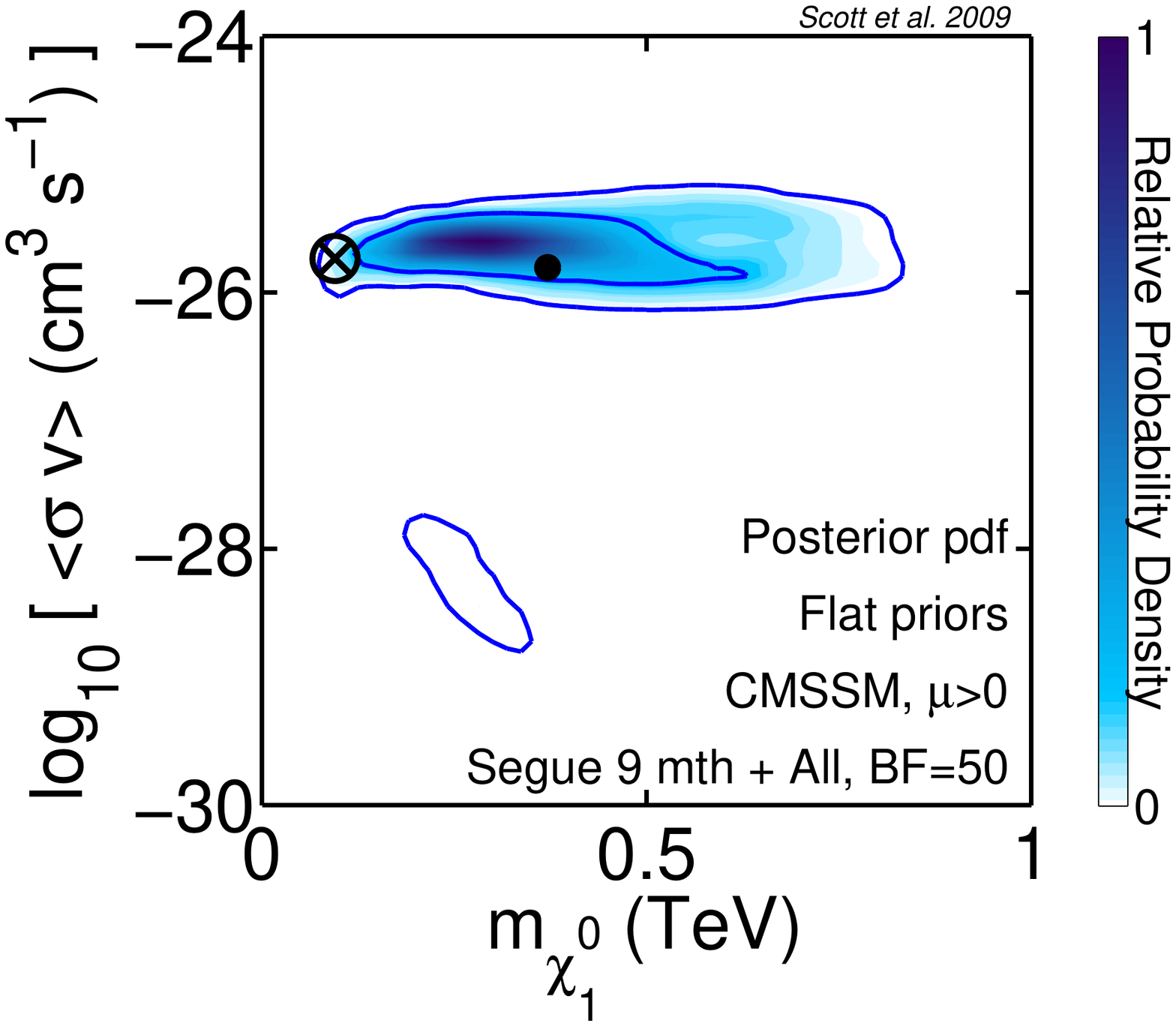}
  \end{minipage}
  \begin{minipage}{0.31\columnwidth}
  \includegraphics[width=\linewidth, trim = 20 170 0 200, clip=true]{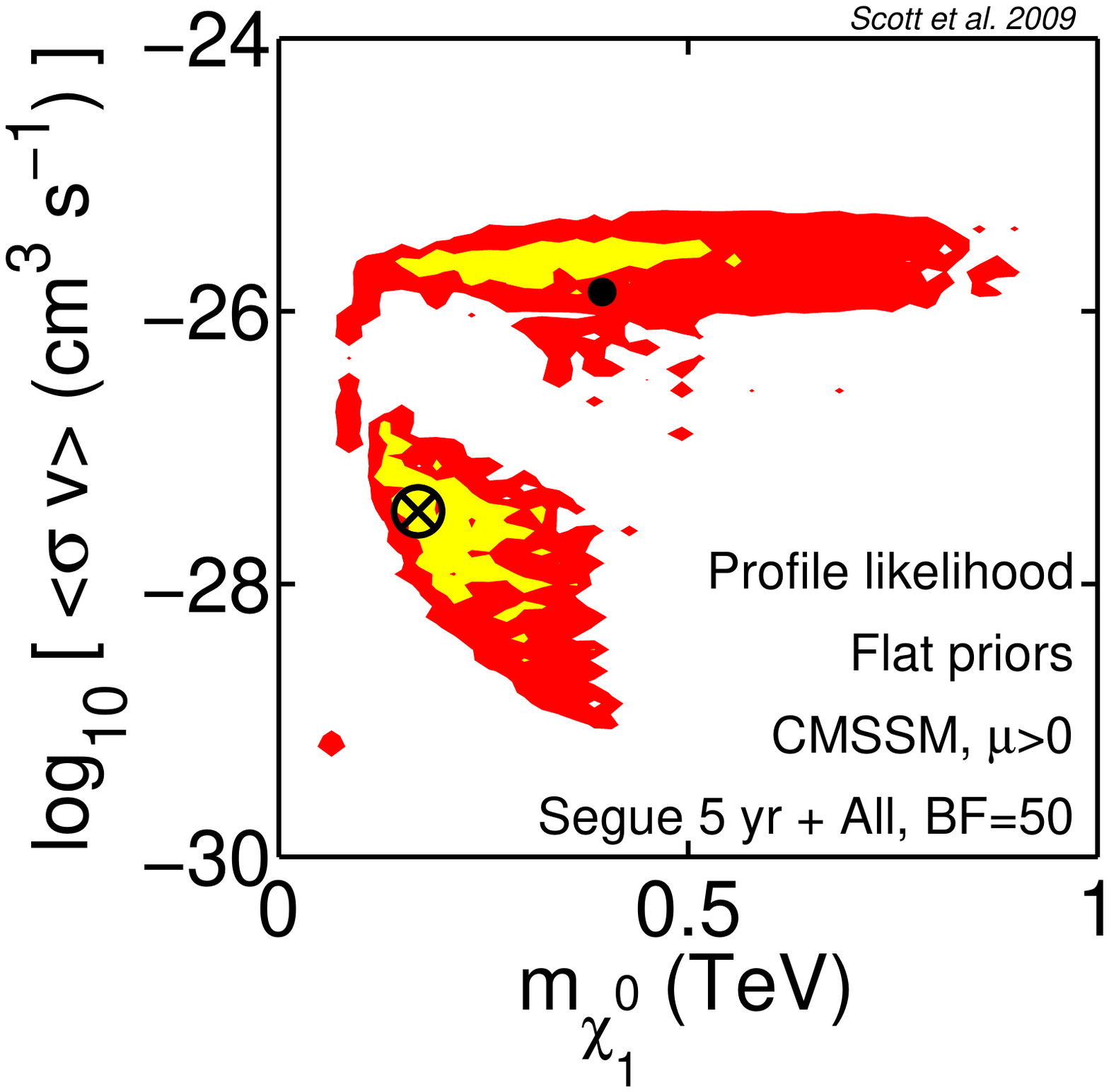}
  \includegraphics[width=\linewidth, trim = 20 170 0 200, clip=true]{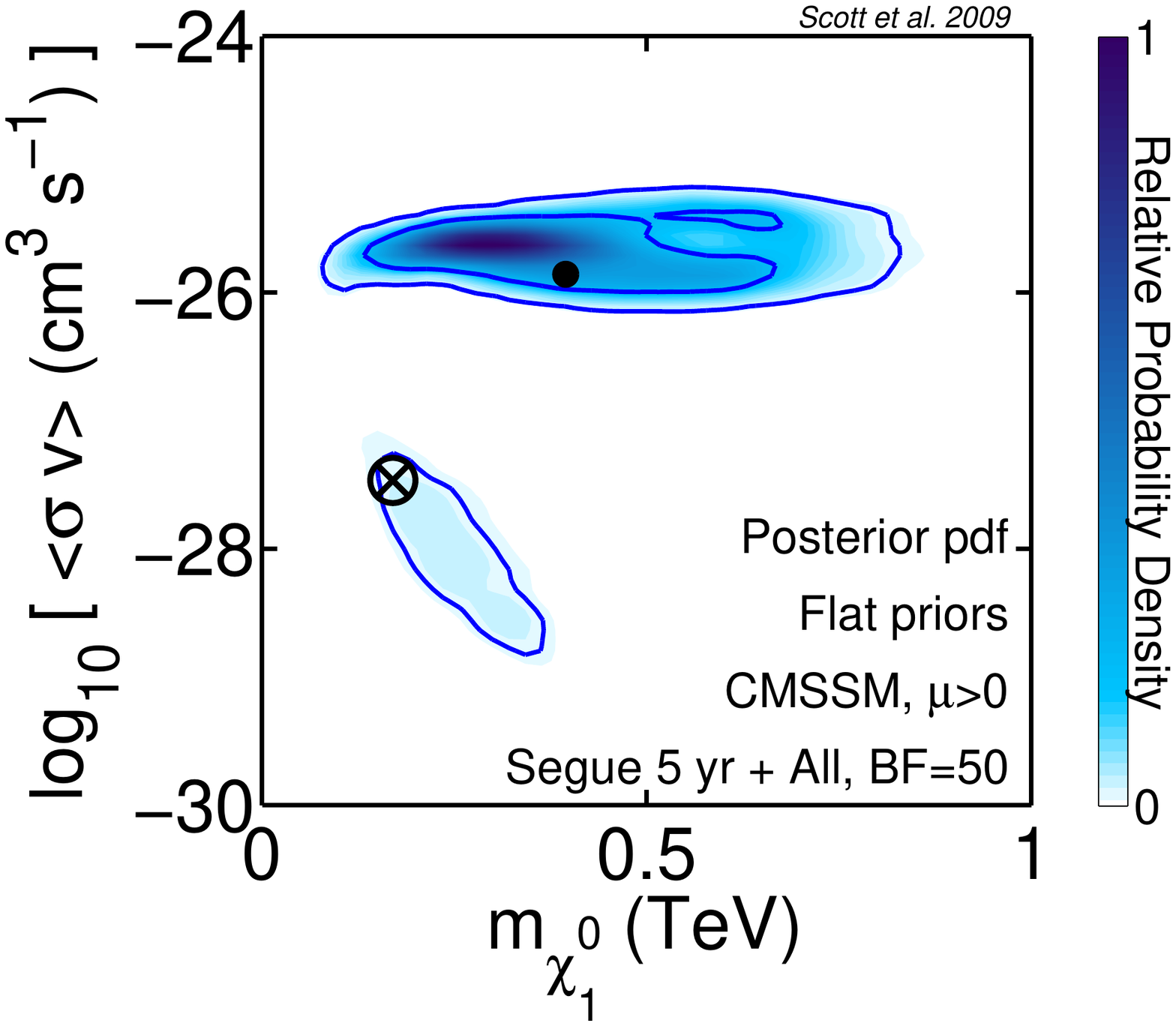}
  \end{minipage}
  \caption{Annihilation cross-sections in the CMSSM which fit all experimental constraints, assuming the neutralino to be the dominant component of dark matter.  Favoured regions are as implied by existing experimental data only (\emph{left}), and with the addition of 9 months of Segue 1 observations by \Fermi (\emph{middle}).  We also show the extrapolated impact of a non-observation of Segue 1 after 5 years (\emph{right}).  Upper plots show profile likelihoods (where yellow and red indicate 68\% and 95\% confidence regions respectively), while lower plots show marginalised posterior PDFs (where solid blue contours give 68\% and 95\% credible regions).  Solid dots indicate posterior means, whereas crosses indicate best-fit points.}
  \label{fig3}
}

\FIGURE[t]{
  \begin{minipage}{0.31\columnwidth}
  \includegraphics[width=\linewidth, trim = 20 170 0 200, clip=true]{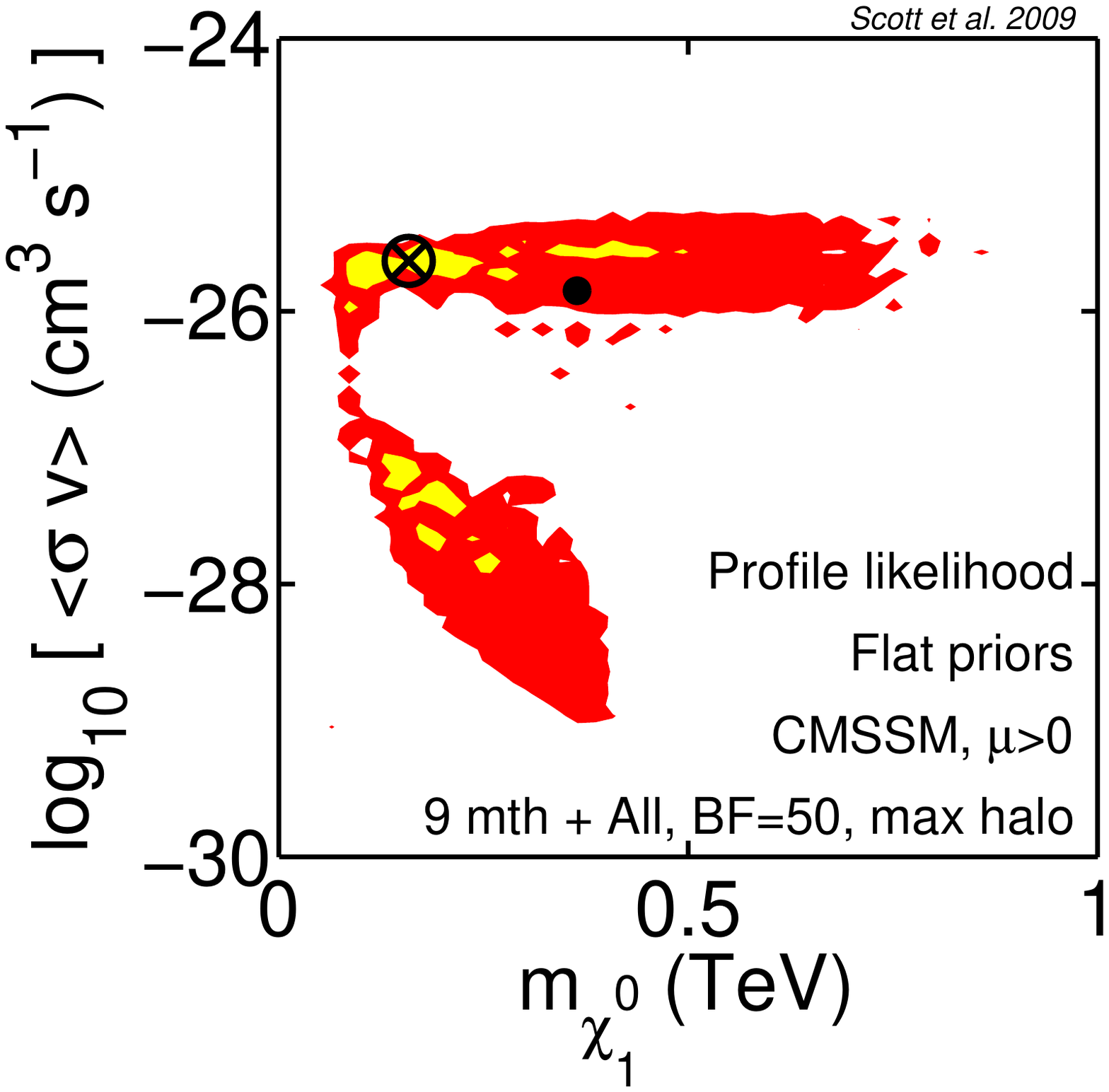}
  \includegraphics[width=\linewidth, trim = 20 170 0 200, clip=true]{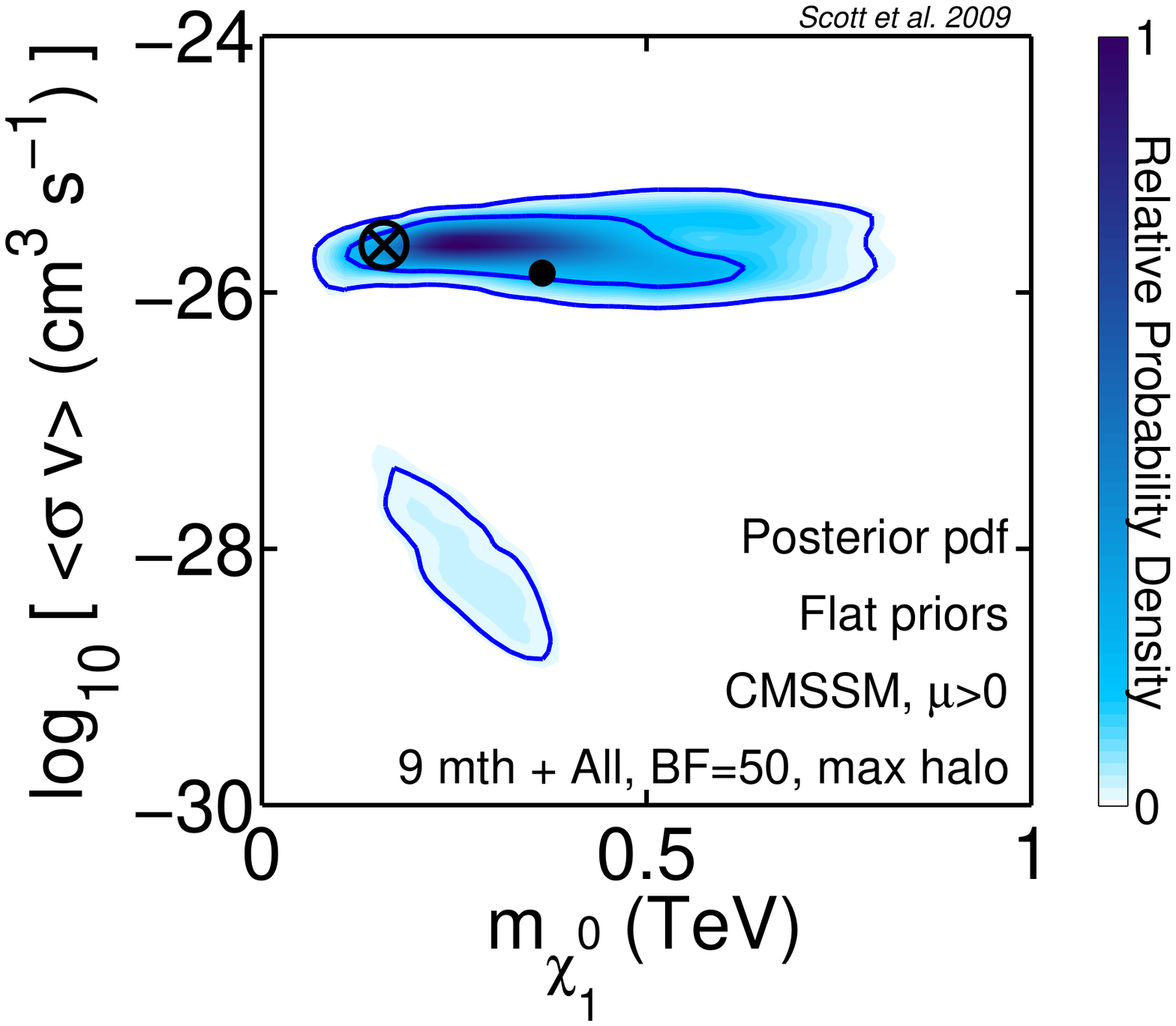}
  \end{minipage}
  \begin{minipage}{0.31\columnwidth}
  \includegraphics[width=\linewidth, trim = 20 170 0 200, clip=true]{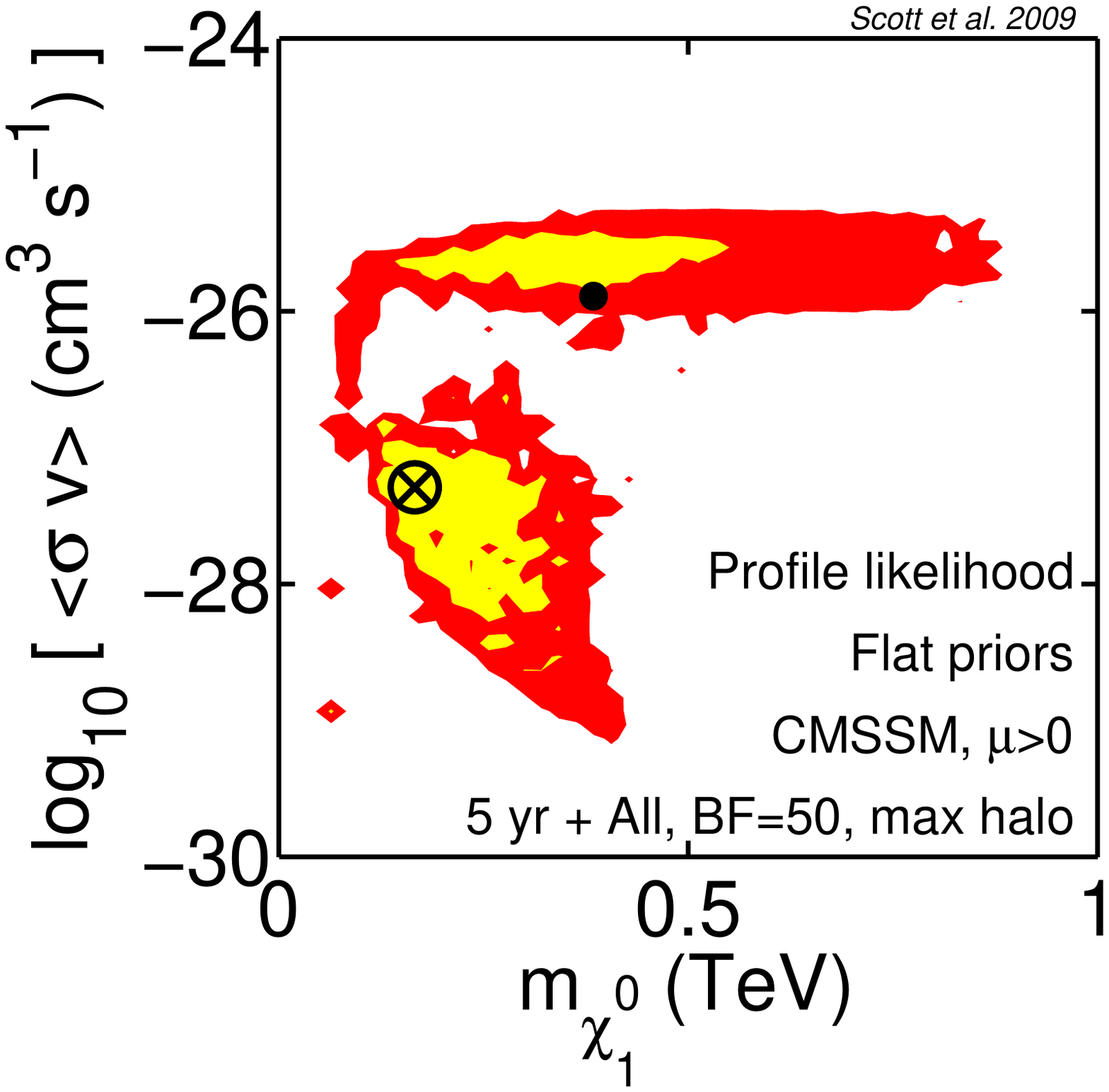}
  \includegraphics[width=\linewidth, trim = 20 170 0 200, clip=true]{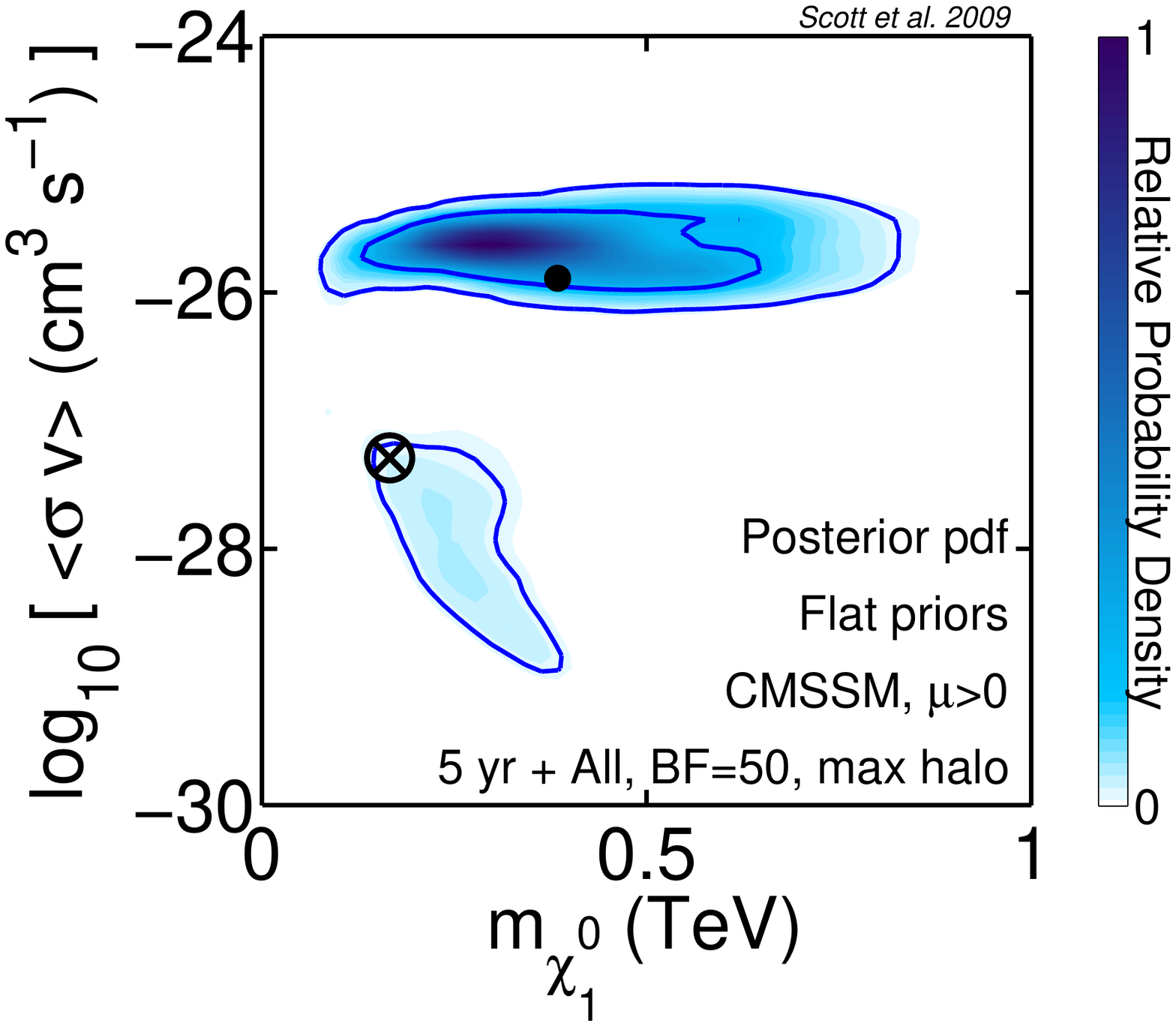}
  \end{minipage}
  \caption{Annihilation cross-sections in the CMSSM which fit all experimental constraints, assuming a `maximally dense' dark matter halo profile for Segue 1.  In this case, the halo scale radius and density were chosen $\sim$$2\sigma$ away from the  best-fit values derived from stellar kinematic data.  Here we again assume the neutralino to be the dominant component of dark matter.  Favoured regions are as implied by 9 months of Segue 1 observations by \Fermi (\emph{left}), and extrapolations to 5 years of data assuming no signal from Segue 1 (\emph{right}).  Shadings and markings are as per Fig.~\protect\ref{fig3}.}
  \label{fig3.5}
}

\FIGURE[t]{
  \begin{minipage}{0.4\columnwidth}
  \centering
  \includegraphics[width=\linewidth, trim = 20 170 0 200, clip=true]{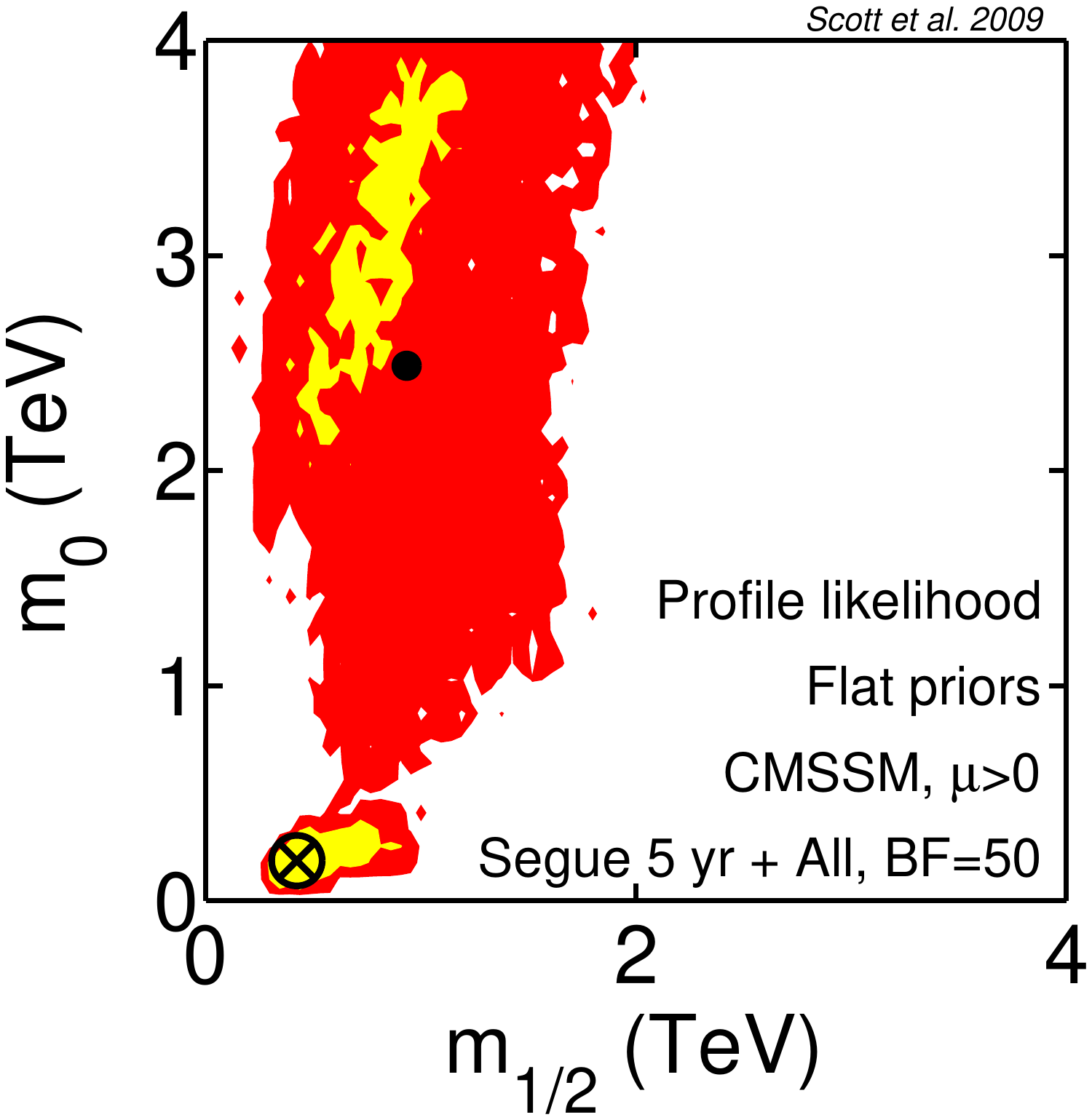}
  \includegraphics[width=\linewidth, trim = 20 170 0 200, clip=true]{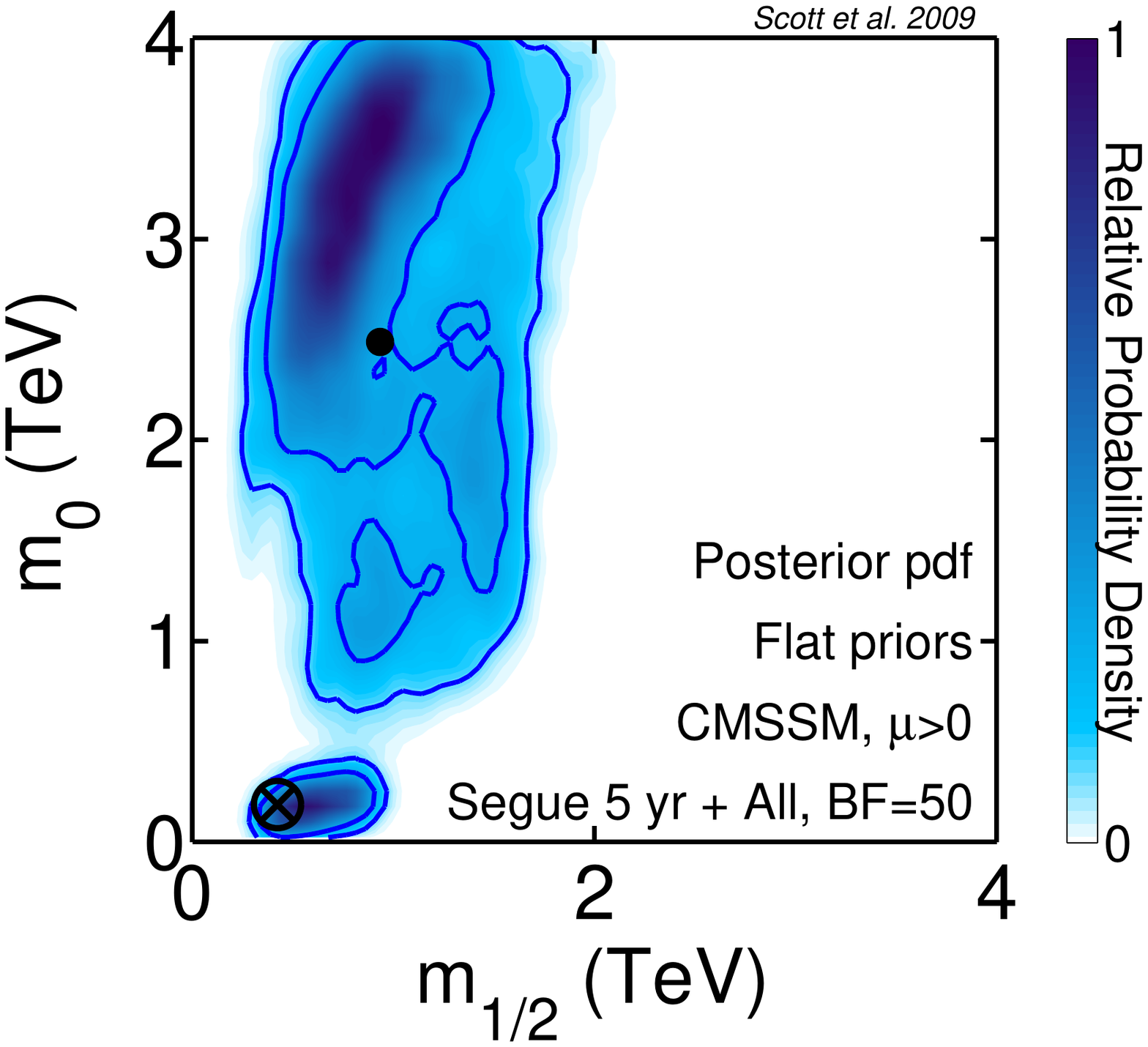}
  \end{minipage}
  \begin{minipage}{0.4\columnwidth}
  \includegraphics[width=\linewidth, trim = 20 170 0 200, clip=true]{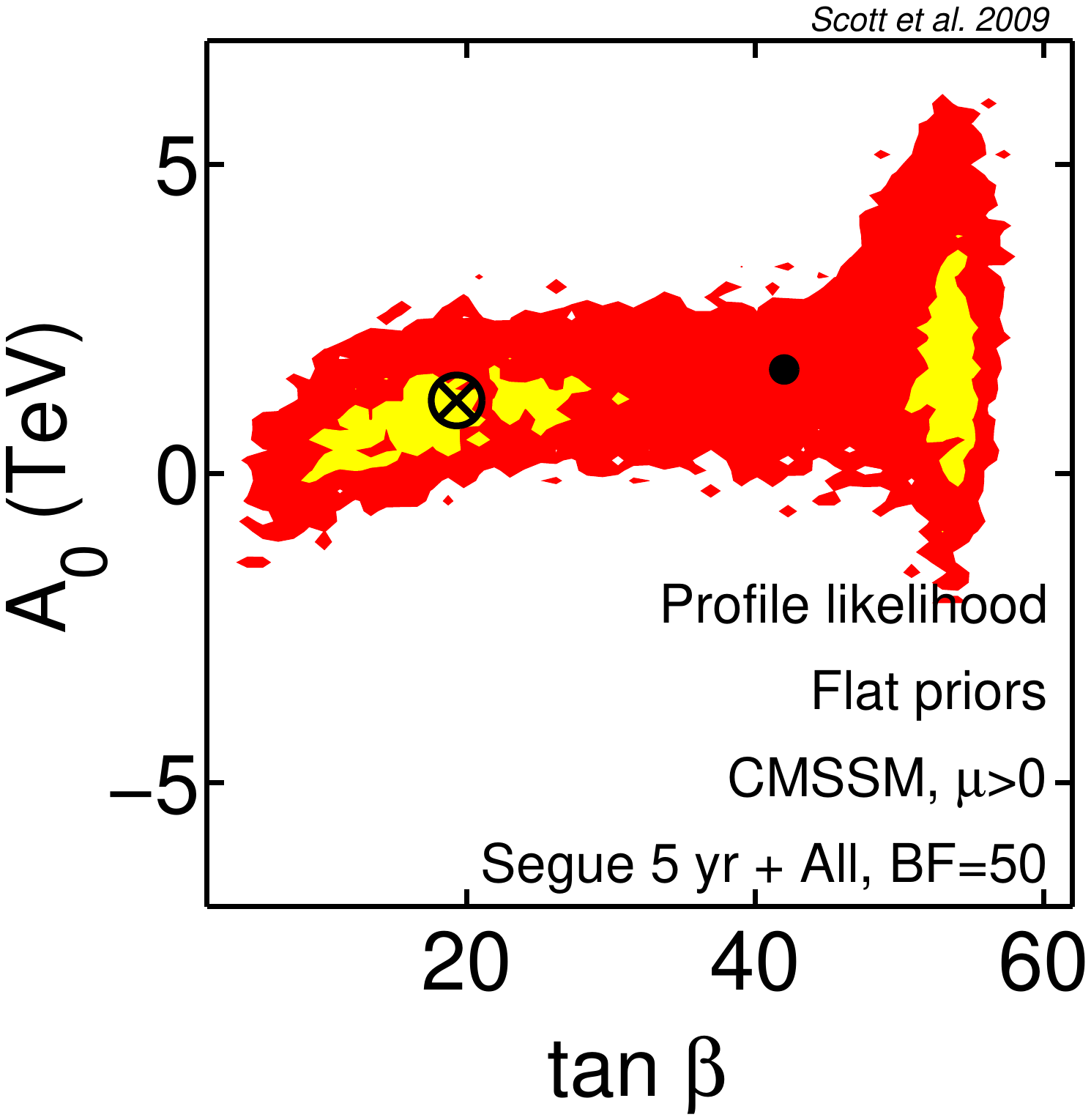}
  \includegraphics[width=\linewidth, trim = 20 170 0 200, clip=true]{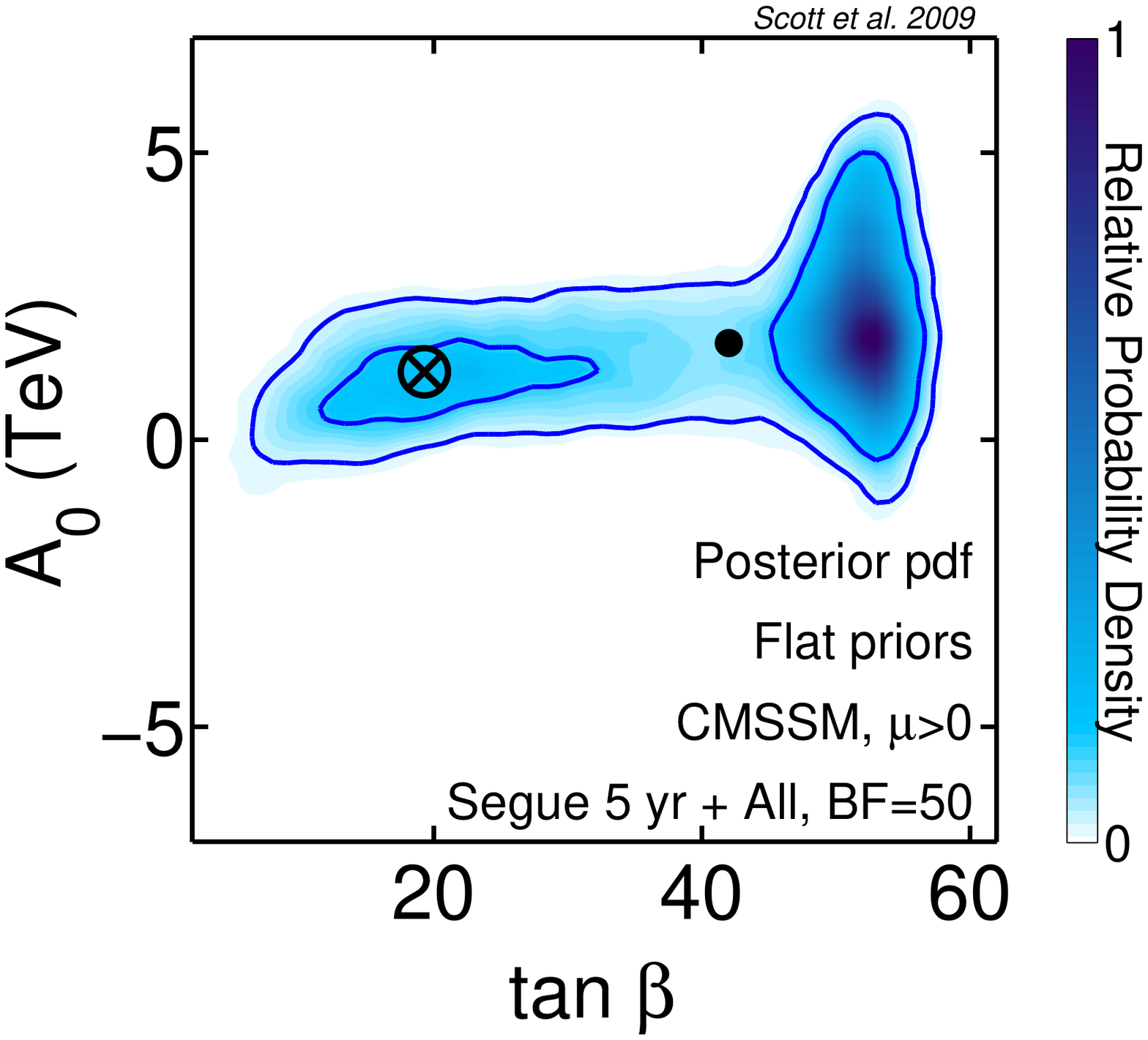}
  \end{minipage}
  \caption{Preferred CMSSM parameter regions including \FLAT observations of Segue 1 and all other observables.  Shadings and markings are as per Fig.~\protect\ref{fig3}.  Preferred regions are very similar whether one considers the existing 9 months of LAT data or extrapolates to 5 years of observations.}
  \label{fig4}
}

\subsection{Global fits}

In Fig.~\ref{fig3} we show the result of including the relic density constraint from the WMAP 5-year data, along with all other experimental bounds.  The effect is to favour models populating two distinct regions: a broad strip around the canonical WIMP annihilation cross-section at $3\times10^{-26}$\,cm$^3$\,s$^{-1}$, and a low-mass region at smaller cross-sections, corresponding to models where stau co-annihilations reduce the relic density to the observed level despite the very low self-annihilation rates.  The models disfavoured by \Fermi observations of Segue 1 in Fig.~\ref{fig2} are here already strongly disfavoured by the relic density constraint, so the additional data from the LAT appears to have little impact upon the preferred cross-sections and masses.  A slight reduction in the profile likelihoods of the lowest mass, highest cross-section corner of the preferred region appears to be present in the extrapolation to 5 years of data.  

The best-fit point is however rather different in the 9-month scan as compared to scans without Fermi data, or with 5 years of mock observations (where we assumed that no excess above background will be seen in 5 years).  In the 9-month scan, the best fit occurs in the focus point region, at a high annihilation cross-section and a low neutralino mass ($\langle\sigma v\rangle = 1.8\times10^{-26}$\,cm$^3$\,s$^{-1}$, $m_\chi=95$\,GeV), whereas the best fits in the other cases are for stau coannihilation models.  This difference appears to be the result of a very small statistical excess above the modelled background in the 9-month data.  Because the corresponding confidence regions are not substantially altered despite the movement of the best fit, the excess would appear to be consistent with observational (statistical) noise.  Given the range of \Fermi's sensitivity, it is thus not at all surprising that the best-fit would appear at this location, falling right on the edge of the instrument's sensitivity.  This point may however be an interesting one to watch as statistics improve.

It is instructive to note the difference in how the co-annihilation region is represented in Fig.~\ref{fig3} by the profile likelihood and the marginalised posterior.  Because the range of CMSSM parameters spanned by the co-annihilation region is quite narrow (i.e. fine-tuned), the total number of points in this region found by the scans is not particularly high, leading to a relatively low posterior PDF.  This is despite the fact that very good fits can be found with a reasonably broad range of neutralino masses and cross-sections in this region, as evidenced by its size in the profile likelihood plots.  In this sense, the Bayesian posterior PDF can be seen to penalise the co-annihilation region to a certain degree for being fine-tuned.  Whether this is a desirable characteristic or not is of course a matter of opinion.  It is, however, important to recognise that such information is only accessible by comparing the posterior PDF and the profile likelihood; the information in their combination is greater than the sum of the parts.

A natural question to ask might be whether more interesting constraints could be obtained from Segue 1 by allowing the neutralino to be a sub-dominant component of dark matter.  Unfortunately, this generally does not add a lot to the discussion when considering constraints from indirect detection with gamma-rays.  Even though the relic density is essentially inversely proportional to the annihilation cross-section, in mixed dark matter scenarios the density of neutralinos in Segue 1 becomes directly proportional to the relic density.  The expected signal is then increased due to the larger annihilation cross-sections permitted by sub-dominant relic densities, but reduced by the reduction in signal due to the reduced galactic densities.  The net result is a reduction in the expected signal, since the flux (Eq.~\ref{fluxeqn}) depends upon the first power of the annihilation cross-section, but the square of the density.  Thus for a decrease in the relic density such that $\Omega_\chi \to \Omega_\chi / X$, the flux is modified as $\Phi \to X/X^2 \Phi = \Phi / X$.  The result is that the favoured cross-sections move to higher values, but the constraints from Segue 1 move even further, providing less constraining power than when the neutralino is assumed to be the only component of dark matter.  This argument of course may not hold for points in the parameter space where the relic density is not strictly inversely proportional to the annihilation cross-section, such as strong co-annihilation or resonant annihilation scenarios.  The former certainly are not probed by the Segue 1 observations in any case, since they lie at very low annihilation cross-sections.  In principle though, highly fine-tuned points in the latter scenario could slightly modify the impact of the Segue constraints in subdominant situations.  As discussed below however, our scans do not uncover a significant number of models where such a mechanism occurs.

In Fig.~\ref{fig3.5} we investigate whether variations in the dark matter profile of Segue 1, within the errors of Martinez et al.~\cite{Martinez09}, might also produce more interesting constraints.  Here we again take an Einasto profile (Eq.~\ref{einasto}), but instead use parameters corresponding to the most dark-matter-rich profile allowed at $\sim$$2\sigma$ ($r_\mathrm{s} = 10$\,pc, $\rho_\mathrm{s}=70$\, GeV\,cm$^{-3}$).  The corresponding constraints on annihilation cross-sections are indeed stronger than in Fig.~\ref{fig3}, but are still largely dominated by the relic density.  This is not surprising, as even though the scale density is a factor of 18 higher in this case, the smaller scale radius means that the higher density occurs at a smaller radius.  In this sense the two parameters are partially degenerate; because \Fermi probes essentially the whole dwarf halo (as Segue 1 should appear almost as a point source), and the total mass of Segue 1 is not substantially altered by the change in halo parameters, the corresponding constraints are not massively improved.  The constraints coming from 9 months of data can be seen to cluster more tightly around the best-fit point at low mass and high cross-section, but not to the point where significant parts of the rest of the parameter space are excluded.  This is consistent with our assertion above that any excess can be explained in terms of statistical fluctuations.

The preferred CMSSM parameter regions including all constraints are shown in Fig.~\ref{fig4}.  Given the marginal impact of Segue 1 observations on scans including the relic density, it is not surprising that the regions are very similar to those shown in \cite{Trotta08}, even when using the extrapolation to 5 years of observations.  The stau co-annihilation region is clearly visible at low $m_0$ and $m_\frac{1}{2}$, separated from the `focus point' region at larger $m_0$.  Scans indicate that both regions are equally well-favoured, though the co-annihilation region tends to return the best-fit point in most cases.  The `bulk' region is mostly disfavoured by relic density and LEP constraints \cite{Allanach06}, but persists at low masses in our scans, overlapping the co-annihilation region in the $m_0$--$m_\frac{1}{2}$ plane. The high-probability region at low $\tan\beta$ in the $A_0$--$\tan\beta$ plane favoured by the co-annihilation region shows up as a much smoother peak in our scans than in some previous works \cite{Ruiz06, Allanach06, Trotta08}.  We suspect that this is due to our use of the upgraded version of \textsf{DarkSUSY} for the relic density calculation.  

The `funnel' region, where resonant annihilation can become important at very low $m_\frac{1}{2}$, does not show up in our scans here.  This is unsurprising, as the nested sampling algorithm is designed to sample according to the total posterior mass, and the linear prior places a very small scanning weight upon such fine-tuned regions at low mass.  Nested sampling routines only find this region when using logarithmic priors on $m_0$ and $m_\frac{1}{2}$ \cite{Trotta08}, though normal MCMC scans can find it a little more easily (e.g. \cite{Ruiz06,Allanach06}).  On the other hand, standard MCMCs and nested sampling implemented with logarithmic priors sample the focus point region less densely, causing them to sometimes miss the highest-likelihood points important for a profile likelihood analysis.  These difficulties are typical consequences using scanning algorithms designed for Bayesian analyses to compute the frequentist profile likelihood; a more promising path for frequentist scans appears to be to use genetic algorithms \cite{Akrami09}.  Using genetic algorithms, it seems possible to find all regions in a prior-independent way, but the ability to effectively map their surroundings and produce reliable confidence regions lags behind other techniques.

Some recent MCMC scans \cite{Buchmueller08, Buchmueller09} have not found large focus-point regions which fit all experimental constraints well, leading the authors to claim that the co-annihilation region is favoured by present data.  In these cases, the reduced likelihood in the focus point region relative to the co-annihilation region was almost entirely due to the fact that it is virtually impossible to produce a good fit to the muon $g-2$ with large values of $m_0$ in the CMSSM.  Using nested sampling with linear priors however, and the physics and likelihood routines within \textsf{SuperBayes}, one can find points in the focus point region where this effect is essentially offset by a correspondingly better fit to other observables \cite{Trotta08}.

\section{Conclusions}
\label{conclusions}

We have incorporated fits to 9 months of \FLAT observations of the dwarf galaxy Segue 1 into explicit global CMSSM parameter scans.  We included gamma-ray lines, internal bremsstrahlung and secondary decay, as well as detailed characterisations of the detector response, its uncertainties and the observed background.  We have also presented scans illustrating the estimated impact of a non-observation of dark matter annihilation in Segue 1 after 5 years of LAT operation.

The LAT data disfavour a small number of physically-viable CMSSM models with low neutralino masses and high annihilation cross-sections, but results depend strongly upon the assumed substructure boost factor in Segue 1.  Such models are already strongly disfavoured by relic density constraints.  Extrapolating to 5 years of operation and assuming the most optimistic boost factor presently allowed by astronomical data, the absence of any annihilation signal from Segue 1 would disfavour all models with cross sections higher than $10^{-25}$\,cm$^3$\,s$^{-1}$, as well as a number at low mass with cross-sections as low as $10^{-26}$\,cm$^3$\,s$^{-1}$.  Even at this level however, the CMSSM models disfavoured by \Fermi would already be essentially excluded by existing data from the microwave background and terrestrial experiments.

\acknowledgments{We are grateful to our colleagues in the \FLAT Dark Matter \& New Physics group for many helpful comments and discussions, and to Riccardo Rando for advice on the LAT IRFs.  We also thank Roberto Trotta for comments on an earlier version of the manuscript.  Through \textsf{DarkSUSY}, \textsf{FLATlib} and \textsf{SuperBayeS} we also drew upon a number of other publicly-available scientific codes, including \textsf{CUBPACK}\cite{cubpack}, \textsf{FeynHiggs}\cite{feynhiggs}, \textsf{SLHALib}\cite{slhalib}, \textsf{SoftSUSY}\cite{softsusy} and \textsf{WCSLIB}\cite{wcslib}.  PS, JC, JE, LB and YA are grateful to the Swedish Research Council (VR) for financial support.  JC is a Royal Swedish Academy of Sciences Research Fellow supported by a grant from the Knut and Alice Wallenberg Foundation.

A number of agencies and institutes have supported both the development and the operation of the LAT as well as scientific data analysis.  These include the National Aeronautics and Space Administration and the Department of Energy in the United States, the Commissariat \`a l'Energie Atomique and the Centre National de la Recherche Scientifique / Institut National de Physique Nucl\'eaire et de Physique des Particules in France, the Agenzia Spaziale Italiana and the Istituto Nazionale di Fisica Nucleare in Italy, the Ministry of Education, Culture, Sports, Science and Technology (MEXT), High Energy Accelerator Research Organisation (KEK) and Japan Aerospace Exploration Agency (JAXA) in Japan, and the K.~A.~Wallenberg Foundation, the Swedish Research Council and the Swedish National Space Board in Sweden.  Additional support for science analysis during the operations phase is gratefully acknowledged from the Istituto Nazionale di Astrofisica in Italy and the and the Centre National d'\'Etudes Spatiales in France.
}

\bibliography{DMbiblio,SUSYbiblio}

\end{document}